\documentclass[conference,compsoc]{IEEEtran}

\usepackage[utf8]{inputenc}
\usepackage[T1]{fontenc}

\usepackage{xcolor}
\usepackage{tikz}
\usetikzlibrary{arrows.meta,positioning,calc,fit}
\usepackage{balance} 
\usepackage{url}
\usepackage{graphicx}
\usepackage{textcomp}
\usepackage{subcaption}
\usepackage[nocompress]{cite}
\usepackage{comment}

\usepackage{makecell}
\usepackage{bm}
\usepackage{amsmath,amsfonts,amsthm,mathtools} 
\usepackage{bbm}
\usepackage[framemethod=TikZ]{mdframed}
\usepackage{hyperref}

\usepackage{booktabs}
\usepackage{multicol}
\usepackage{multirow}
\usepackage{colortbl}
\usepackage{tcolorbox}
\tcbuselibrary{breakable,skins}

\usepackage{soul}

\usepackage{wrapfig}
\usepackage{xspace}
\usepackage{bbding}
\usepackage{pifont}

\usepackage{wasysym}
\usepackage{threeparttable}

\usepackage[linesnumbered,ruled,vlined]{algorithm2e} 
\usepackage{cleveref}

\usepackage{threeparttable}

\usepackage{acronym}
\usepackage{xltabular}

\usepackage{marvosym}

\usepackage{CJK}

\theoremstyle{definition}
\newtheorem{definition}{Definition}

\newenvironment{packeditemize}{
\begin{list}{$\bullet$}{
\setlength{\labelwidth}{8pt}
\setlength{\itemsep}{2pt}
\setlength{\leftmargin}{\labelwidth}
\addtolength{\leftmargin}{\labelsep}
\setlength{\parindent}{0pt}
\setlength{\listparindent}{\parindent}
\setlength{\parsep}{0pt}
\setlength{\topsep}{2pt}}}{\end{list}}

\newenvironment{packedenumerate}{
\begin{list}{\arabic{enumi}.}{
\usecounter{enumi}
\setlength{\labelwidth}{10pt}
\setlength{\itemsep}{2pt}
\setlength{\leftmargin}{\labelwidth}
\addtolength{\leftmargin}{\labelsep}
\setlength{\parindent}{0pt}
\setlength{\listparindent}{\parindent}
\setlength{\parsep}{0pt}
\setlength{\topsep}{2pt}}}
{\end{list}}

\acrodef{llm}[LLM]{Large Language Model}
\acrodef{rag}[RAG]{Retrieval-Augmented Generation}
\acrodef{rvd}[RVD]{recurring vulnerability detection}
\acrodef{sast}[SAST]{Static Application Security Testing}

\newcommand{\deepseek}{\textsc{DeepSeek}\xspace}
\newcommand{\sysname}{\textsc{InfraScope}\xspace}

\newcommand{\codeql}{\textsc{CodeQL}\xspace}

\newcommand{\refvul}{\ensuremath{v^{\star}}\xspace}
\newcommand{\varvul}{\ensuremath{v}\xspace}
\newcommand{\refrepo}{\ensuremath{S^{\star}}\xspace}
\newcommand{\targetrepo}{\ensuremath{S}\xspace}

\newcommand{\reposem}[1]{\ensuremath{\Sigma_R(#1)}}
\newcommand{\witness}[1]{\ensuremath{\pi(#1)}}

\newcommand{\refvulnsem}{\ensuremath{\Sigma_V(\refvul)}\xspace}
\newcommand{\varvulnsem}{\ensuremath{\Sigma_V(\varvul)}\xspace}
\newcommand{\refreposem}{\ensuremath{\Sigma_R(\refrepo)}\xspace}
\newcommand{\targetreposem}{\ensuremath{\Sigma_R(\targetrepo)}\xspace}

\newcommand{\bheading}[1]{\noindent\textbf{#1}\ }

\newtcolorbox[auto counter]{findingbox}[1][]{
  enhanced jigsaw,
  breakable,
  colback=black!2,
  colframe=black!65,
  boxrule=0.6pt,
  arc=1pt,
  boxsep=0pt,
  left=4pt,
  right=4pt,
  top=4pt,
  bottom=4pt,
  before skip=4pt,
  after skip=4pt,
  fontupper=\small,
  #1
}

\newcommand{\findingtext}[1]{\noindent\textbf{Finding~\thetcbcounter:} #1}

\newcommand{\etal}{\emph{et al.}\xspace}
\newcommand{\etc}{\emph{etc}\xspace}
\newcommand{\ie}{\emph{i.e.}\xspace}
\newcommand{\eg}{\emph{e.g.}\xspace}

\newcommand{\symyes}{\ensuremath{\bullet}}
\newcommand{\sympart}{\ensuremath{\odot}}
\newcommand{\symno}{\ensuremath{\circ}}

\begin{document}

\title{Hunting Vulnerability Variants in AI Infra: Measurement and Reference-Driven Detection}

\author{
\IEEEauthorblockN{
Tian Dong\IEEEauthorrefmark{1},
Yanjun Chen\IEEEauthorrefmark{2},
Shoufeng Zhang\IEEEauthorrefmark{1},
Huaien Zhang\IEEEauthorrefmark{1}\IEEEauthorrefmark{4},\\
Yunlong Lyu\IEEEauthorrefmark{1},
Keke Lian\IEEEauthorrefmark{2},
Dong Zhang\IEEEauthorrefmark{2},
Shaofeng Li\IEEEauthorrefmark{3},
Hao Chen\IEEEauthorrefmark{1}
}
\IEEEauthorblockA{\IEEEauthorrefmark{1}The University of Hong Kong}
\IEEEauthorblockA{\IEEEauthorrefmark{2}Tencent}
\IEEEauthorblockA{\IEEEauthorrefmark{3}Southeast University}
}

\maketitle
\begingroup
\renewcommand{\thefootnote}{\fnsymbol{footnote}}
\footnotetext[4]{Corresponding author.}
\endgroup

\begin{abstract}

AI infra has become a shared execution layer for model training, deployment, and agent orchestration.
Because many projects reimplement similar model-centric workflows, a vulnerability disclosed in one repository can recur as a variant in another repository with a related design.
Yet the prevalence and detectability of these variants remain poorly understood.
This paper presents a measurement study of vulnerability variants in AI infra. Analyzing 688 GitHub repositories and 251 publicly disclosed vulnerabilities, we find that AI infra projects frequently share overlapping functionality and recurrent vulnerable patterns, creating a concrete basis for cross-repository variants.
Building on this finding, we study how to automatically identify such variants from known disclosures. We propose \sysname, a reference-driven multi-agent framework that extracts transferable vulnerability semantics from known cases and uses them to locate and validate variants in new repositories. Evaluating \sysname on 20 real-world AI infra repositories, we uncover over 20 vulnerabilities, including 11 acknowledged cases and 4 cases that have been assigned CVEs so far.
\end{abstract}

\IEEEpeerreviewmaketitle

\section{Introduction}

Once an AI infra vulnerability is disclosed, the same failure mode may already exist elsewhere.
Modern AI infra is now on the execution path of model post-training~\cite{zheng-etal-2024-llamafactory}, serving~\cite{vllm_docs}, and agent orchestration~\cite{yao2023react}.
In these workflows, a developer or end user may upload an adapter, point to a remote checkpoint, or configure a tool through a Web UI, CLI, or serving API.
The infrastructure layer~\cite{infraforagent_tmlr25} then fetches external data, deserializes model state, launches commands, and connects models to downstream tools or data sources.
This pairing of attacker-controlled inputs with privileged operations makes shared AI infra software a natural attack surface.
A public LlamaFactory disclosure (70k+ GitHub stars) illustrates the risk: user-controlled model or adapter paths can reach unsafe model-loading logic and cause remote code execution in repositories that implement related post-training workflows~\cite{cve2025_53002_nvd}.
As codebase-scale autonomous bug discovery becomes increasingly practical (\eg, Mythos~\cite{mythos2026}), each disclosure creates an urgent post-disclosure auditing problem: maintainers must determine whether the same trigger semantics recur in related repositories before attackers scale the search.

This audit cannot be solved by code similarity alone.
AI infra projects often realize the same functional context through different wrappers, module boundaries, and deployment assumptions.
Existing recurring vulnerability detectors work well when transfer follows code reuse~\cite{kim2017vuddy,woo2022movery,woo2023v1scan}, patch lineage~\cite{li2017securitypatches,woo2022movery,huang2024vmud}, or stable signatures~\cite{feng2024fire,xiao2024accurate,yang2026iotbec}.
These assumptions do not directly cover a disclosed trigger mechanism that reappears in semantically similar forms across independently developed AI infra projects.
Repository-scale auditors and \ac{llm}-assisted analyzers broaden codebase exploration through agentic search~\cite{repoaudit2025} or neuro-symbolic reasoning~\cite{li2025iris}.
However, they are usually organized around general bug classes rather than preserving the semantic features of a known reference vulnerability across related repositories.
Project-scale studies further show that \ac{llm}-based repository analysis remains constrained by shallow interprocedural reasoning~\cite{li2026projscale} and bounded context~\cite{liu2023lostmiddle}.
These limitations call for an audit procedure that starts from a known vulnerability and searches related AI infra projects for semantically equivalent variants.

\bheading{Motivation.}
Before building such a procedure, we first examine whether AI infra provides a real empirical basis for cross-repository vulnerability variants.
This leads to three research questions:
\begin{packedenumerate}
    \item Do public AI infra repositories share enough functional context for cross-repository transfer to be meaningful?
    \item Given a public AI infra vulnerability, can its trigger mechanism appear as semantically equivalent variants in repositories with similar functional context?
    \item If so, can a public disclosure be turned into a practical auditing procedure that identifies such variants in related repositories?
\end{packedenumerate}

\bheading{Measurement.}
We answer the first two questions through a measurement of the public AI infra ecosystem (\Cref{sec:landscape}).
We crawl and analyze 688 AI infra repositories and 251 public vulnerabilities, including 232 disclosures with trigger details sufficient for variant analysis.
The results show that AI infra is rapidly expanding, concentrated in a few major capability families, and repeatedly built from overlapping functional modules.
More importantly, public disclosures expose recurring trigger mechanisms under shared functional contexts, not merely isolated project-specific failures.
This evidence motivates post-disclosure auditing across related repositories.

\bheading{\sysname.}
We therefore study reference-driven auditing for vulnerability variants (\Cref{sec:motivation,sec:method}).
We call the disclosed case a \emph{reference vulnerability} and a semantically equivalent case in another repository a \emph{vulnerability variant}.
The core idea is to make manual variant hunting practical by placing tool-augmented LLMs inside a controlled auditing harness.
Directly applying agents with conventional software development tools faces three challenges.
First, a reference vulnerability gives only partial guidance; an unconstrained agent can spend its budget on irrelevant modules or false candidates.
Second, current LLMs have limited context windows~\cite{repoaudit2025,liu2023lostmiddle}, making it difficult to sustain a long inspection and preserve relevant memory across sessions.
Third, hallucination can produce nonexistent candidates or unsupported conclusions that are hard to filter without manual verification.

We propose \sysname, a multi-agent framework for identifying vulnerability variants in AI infra.
\sysname addresses these challenges through a reference-driven auditing harness that coordinates specialized agents for semantic modeling, localization, and verification.
First, a semantic modeling agent extracts vulnerability features (\eg, propagation) from the reference and models the target repository as inspection-oriented functional modules with module-level dependencies.
Second, an inspection agent uses localization-aware state management to prioritize target modules with matching AI infra functionality and externalize compact audit state, such as inspected files and rejected hypotheses, across bounded sessions.
Third, a target-side verification agent combines repository code facts with automated PoC generation to check exploitability before reporting.
The workflow assumes existing vulnerabilities and repository code, making post-disclosure auditing practical without private telemetry or privileged deployment access while keeping final claims grounded in repository facts.

We implement \sysname in over 29K lines of code and evaluate it with 8 reference vulnerabilities and 20 target repositories.
In the benchmark, \sysname reports 31 candidates, including 24 true positives, and achieves the highest precision and accuracy among the compared tools.
It consumes over 7$\times$ fewer tokens than Claude Code under the same references while detecting only 4\% fewer true positives.
Across the benchmark and an OpenClaw case study, \sysname uncovers over 20 zero-day vulnerabilities, including 11 acknowledged cases and 4 assigned CVEs so far.

\bheading{Contributions.}
This paper makes the following contributions:
\begin{packeditemize}
    \item We conduct a measurement across 688 AI infra repositories and 251 related public vulnerabilities, identifying recurring overlaps in functionality and vulnerability patterns.
    \item We formalize reference-driven vulnerability variant auditing, where the known vulnerability defines the trigger semantics to preserve and functional context guides localization.
    \item We design and implement \sysname, which coordinates three agents on semantics modeling, functional-context localization, and target-repository verification.
    \item We evaluate \sysname on 8 reference vulnerabilities and 20 target repositories, finding over 20 zero-day vulnerabilities, including 11 acknowledged cases and 4 assigned CVEs so far.
\end{packeditemize}

\begin{figure*}[t]
    \centering
    \includegraphics[width=\textwidth]{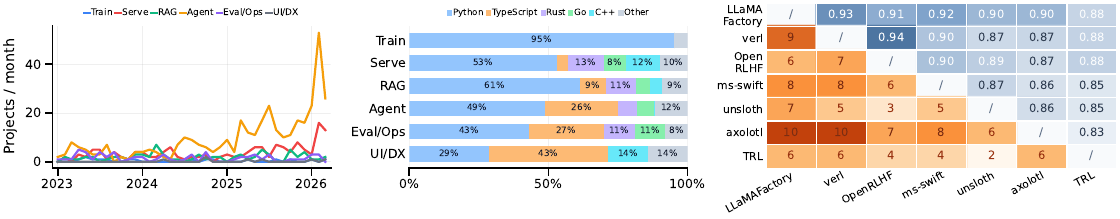}
    \begin{subcaptiongroup}
        \phantomsubcaption\label{fig:ai-infra-landscape}
        \phantomsubcaption\label{fig:ai-infra-language-mix}
        \phantomsubcaption\label{fig:training-cluster-similarity}
    \end{subcaptiongroup}
    \caption{AI infra measurement summary. {\normalfont (a) monthly project creation by capability family since 2023, (b) primary-language composition within each family, and (c) project-level closeness among the seven analyzed training frameworks: README embedding similarity in the upper triangle and shared training stack features in the lower triangle.}}
    \label{fig:ai-infra-measurement-triptych}
\end{figure*}

\section{AI Infra Ecosystem}
\label{sec:landscape}
In this section, we measure recent AI infra ecosystem growth, examine cross-project functional similarity, and test whether public vulnerabilities expose similar trigger mechanisms across related functional contexts.

We use \emph{AI infra} to denote a reusable implementation layer that lets developers build and operate AI systems.
For example, an agent orchestration framework (\eg, \texttt{openclaw}) can be integrated with a finetuning framework (\eg, \texttt{transformers}) to build a personalized assistant.
AI infra typically includes one or more capabilities such as post-training, serving and inference, data retrieval, agent orchestration, and user interfaces, \etc.
We exclude generic repositories such as tutorials or prompting guides unless they provide AI-specific reusable infrastructure functionality.
To simplify, we manually assign each repository a primary capability label in the measurement and detail the module-role taxonomy in \Cref{sec:method}.

\bheading{Measurement setup.}
We first present a summarized measurement setup and refer the reader to Appendix~\ref{sec:appendix-training-cluster-measurement} for more details.
We construct a dataset of 688 AI infra repositories through the GitHub Search REST API using topic and keyword queries.
For growth, we focus on the 676 repositories created on or after January 1, 2023. Repository-level summaries use all 688 repositories.
For each repository, we retain the GitHub information needed for the measurement (\eg, description and primary language).
We exclude repositories that only mention AI terms but contain auxiliary material or end-user applications rather than reusable infrastructure.

\bheading{Growth tendency.}
We begin with project counts and programming language usage in the collected AI infra repositories.
\Cref{fig:ai-infra-landscape} shows monthly counts of newly created AI infra projects by capability family in our collection.
The ecosystem continues to expand across multiple capability families.
Specifically, within the 2023+ discovery window, the annual counts of newly created projects increase from 124 in 2023 to 142 in 2024 and 262 in 2025, while the first quarter of 2026 already contributes 148 projects.
Notably, the agent orchestration is the largest family in our collection with 373 projects, but serving and inference, RAG and data, and evaluation and observability also contribute substantial volume, with 149, 75, and 63 projects, respectively.

\Cref{fig:ai-infra-landscape} also shows a peak in February 2026, when 73 new repositories appear overall and 53 of them fall into agent orchestration and the full monthly cohort spans 72 distinct owners.
To test whether OpenClaw imitation explains this peak, we conduct a follow-up OpenClaw-centered similarity check on those 53 agent orchestration repositories.
In total, we identify 36 OpenClaw-like cases, 9 partially similar cases, and 8 dissimilar cases, suggesting broader family-level reuse.
To distinguish creation bursts from repository liveness, we also record each repository's latest push and study how recently the collected repositories were updated.
We find that 428 repositories were last pushed in March 2026, and this concentration is not explained by OpenClaw-similar repositories, which account for only 6 of the 428 cases.
Together with the creation trend, the latest push trend indicates that AI infra is both expanding and recently maintained at scale.

We next examine whether this active ecosystem is concentrated in a few language and capability families, which would support the feasibility of \sysname.

\bheading{Language and type concentration.}
In \Cref{fig:ai-infra-language-mix}, we investigate primary-language distributions across the same capability families using the primary language reported by GitHub.
The remaining share is grouped into \emph{Other}, such as JavaScript in capability families where those languages are present but not separately labeled.
As expected, Python is the largest language overall, appearing in 356 of 688 projects.
It remains the largest language in agent orchestration, serving and inference, and RAG and data, while TypeScript, Go, Rust, and C++ are also widely used to accelerate programs.

\Cref{tab:ai-infra-type-share} complements this within-family view with family-level repository and star shares.
Agent orchestration also dominates star share, while serving and inference and RAG and data form a second tier by project volume.
The platform and training family remains smaller by project count but still captures a non-trivial star share, indicating influence concentrated in a smaller set of repositories.
The language and family shares show concentration in several major capability families, but family labels alone do not establish transferable functionality.
We therefore inspect whether related repositories expose similar implementation features.

\begin{table}[t]
    \centering
    \caption{Repository and star shares of AI infra capability families in the dataset.}
    \resizebox{\linewidth}{!}{
    \begin{tabular}{lrrrr}
        \toprule
        \textbf{AI infra Type} & \textbf{Projects} & \textbf{Project Share} & \textbf{Stars} & \textbf{Star Share} \\
        \midrule
        Agent orchestration & 373 & 54.2\% & 3{,}114{,}853 & 61.6\% \\
        Serving and inference & 149 & 21.7\% & 511{,}233 & 10.1\% \\
        RAG and data & 75 & 10.9\% & 699{,}624 & 13.8\% \\
        Evaluation and observability & 63 & 9.2\% & 285{,}164 & 5.6\% \\
        Platform and training & 21 & 3.1\% & 437{,}192 & 8.6\% \\
        UI and workflows & 7 & 1.0\% & 10{,}400 & 0.2\% \\
        \bottomrule
    \end{tabular}
    }
    \label{tab:ai-infra-type-share}
\end{table}

To further examine functional similarity between projects, we take a closer look at training frameworks, a type of AI infra with repeated workflows and a common programming language (Python).

\bheading{Functionality closeness.}
We select the seven representative training frameworks~\cite{zheng-etal-2024-llamafactory,sheng2024hybridflow}, which collectively exceed 200k GitHub stars as of March 2026.
We represent each framework with 12 binary structural and workflow features such as distributed training and inference augmentation.
We also compute README-level semantic similarity using normalized embeddings produced by \path{bge-large-en-v1.5}~\cite{bge_embedding}.
\Cref{fig:training-cluster-similarity} shows two complementary measurements.
Entries in the upper triangle report README embedding similarity, whereas entries in the lower triangle count how many of the 12 structural and workflow features are shared by each framework pair.
The seven frameworks are not near-duplicate projects, but their README semantics and shared workflow features still form visible similarity clusters.
The median off-diagonal embedding similarity is 0.88 (IQR 0.04), and the strongest pair reaches 0.94 for verl and OpenRLHF, followed by 0.93 for verl and LLaMA-Factory and 0.92 for LLaMA-Factory and ms-swift.

\begin{findingbox}
\findingtext{AI infra is a rapidly expanding, Python-centered ecosystem with overlapping functional modules. This overlap supplies the functional substrate required for similarity-guided variant transfer across modules and projects.}
\end{findingbox}

\bheading{Trigger-level similarity in public disclosures.}
The remaining question is whether shared AI infra functionality can lead to vulnerability variants, which share a similar functional context, trigger scenario, and vulnerability family.
We curate a dataset of publicly disclosed vulnerabilities from public advisory databases~\cite{nvd_general,osv_dev,github_advisory_database}, vendor writeups and huntr reports~\cite{huntr_platform}.

The main dataset covers disclosures since January 1, 2024 and keeps only records that map back to reusable AI infra repositories or packages.
To avoid treating sparse metadata as support, we require public trigger details to investigate whether similar functionality carries similar vulnerability mechanisms, not merely whether projects share broad CWE labels.
After manually deduplicating multiple sources for the same disclosure and excluding non-AI infra records, the main dataset contains 597 source records covering 251 public vulnerabilities across 18 projects.
Among them, 232 have public trigger details.

\begin{figure}[t]
    \centering
    \includegraphics[width=\columnwidth]{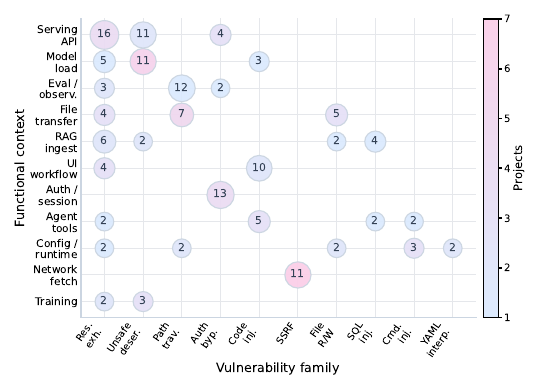}
    \caption{Similar vulnerability patterns in AI infra disclosures. Bubbles aggregate functional-context and vulnerability-family pairs supported by public trigger details. Number and area encode deduplicated vulnerability count, and color encodes affected projects.}
    \label{fig:ai-infra-vuln-context-map}
\end{figure}

\Cref{fig:ai-infra-vuln-context-map} summarizes these patterns with public trigger details after grouping rows by context and vulnerability family.
The strongest cross-project patterns include serving API resource exhaustion (16 deduplicated vulnerabilities across 4 projects), access-control bypasses (13 across 4 projects), attacker-controlled URL fetches leading to SSRF (11 across 7 projects), and model-loading deserialization bugs (9 across 5 projects).
\begin{findingbox}
\findingtext{Public AI infra disclosures reveal similar trigger mechanisms under shared functional contexts across projects, rather than only isolated project-specific failures. This evidence justifies functionality-guided localization, while reportable variants still require semantic consistency in the target code.}
\end{findingbox}

\section{Vulnerability Variants in AI Infra}
\label{sec:motivation}

In this section, we first show a motivating example of vulnerability variants across two training repositories, then derive the design challenges and finally formalize the detection problem.

\begin{figure*}[t]
    \centering
    \includegraphics[width=\textwidth]{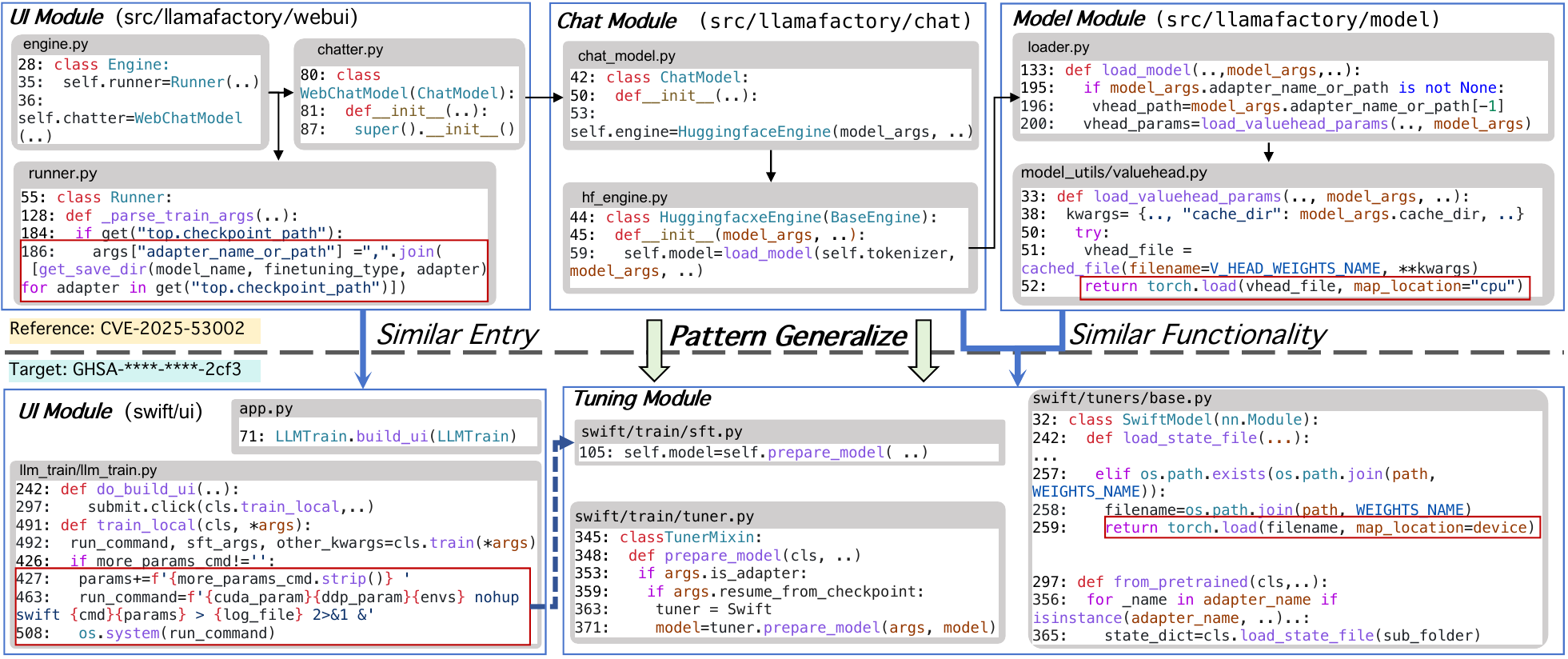}
    \caption{Cross-repository motivating example for vulnerability variants in AI infra found by \sysname. A disclosed LLaMA-Factory RCE (CVE-2025-53002) and an ms-swift advisory (GHSA-****-****-2cf3) share a similar source in the UI module, where a user-controlled adapter path reaches the unsafe loading sink \texttt{torch.load} through parameter passing and command invocation ({\color{blue}{dashed arrow}}), respectively.}
    \label{fig:motivating-examples}
\end{figure*}

\subsection{Motivating Example}
Vulnerability variants are not limited to directly copied flaws; they preserve the semantics of the reference vulnerability despite code changes such as refactoring and API migration across AI infra projects.
They can exist across different modules in the same AI infra project or across similar repositories.
For instance, more than 17 SQL injection variants appear in different database modules of the agent orchestration project \textit{llama\_index} (see \Cref{fig:within-project-variant}).
We next present a cross-repository example of vulnerability variants as shown in \Cref{fig:motivating-examples}.
In CVE-2025-53002, the affected \texttt{LLaMA-Factory} code receives a user-uploaded adapter path (\ie, source).
The path can point to malicious serialized data from an attacker-controlled remote source (\eg, a Hugging Face model).
Then, the vulnerable code directly deserializes the adapter with \texttt{torch.load} (\ie, sink), which can lead to remote code execution (RCE) if the attacker crafts a malicious payload.

The same question arises for \texttt{ms-swift}, which implements a comparable post-training capability.
The target contains related UI logic and model-loading modules, but the propagation path differs.
In \texttt{ms-swift}, model loading relies not on direct function calls but on command execution through \texttt{swift --params} in \texttt{llm\_train.py}.
A call graph-based approach that searches for a direct relation from the source to sink would not model this command-mediated propagation and can miss the variant.

\sysname found the vulnerability variant GHSA-****-****-2cf3 by transferring the trigger semantics of CVE-2025-53002.
This case starts from a WebUI or training entry and propagates the user-provided adapter model path into the executed command.
The command executed through \texttt{os.system} eventually reaches \texttt{torch.load} without a secure deserialization check~\cite{cve2025_53002_nvd}.
Although the two vulnerabilities expose the same deserialization pattern (a UI-triggered training path reaches tuning and model-loading logic that deserializes adapter state with \texttt{torch.load}), they manifest in different affected modules due to distinct implementations.

\subsection{Challenges}
This example motivates our reference-driven formulation: the transferred object is trigger semantics, not a hosting boundary, exact call graph, or coarse CWE\@.
Therefore, we require three transfer cues.
\emph{Similar entry} identifies user-facing training functionality rather than identical code.
\emph{Pattern generalization} abstracts the disclosed bug into an attacker-controlled asset path, cross-module propagation, and missing guard.
\emph{Similar functionality} localizes target modules that implement similar behavior.
LLM-assisted vulnerability analyzers either augment static analysis with taint specifications, code property slices, or constraints~\cite{li2025iris,lekssays2025llmxcpg,li2025vulsolver}, or use repository-level agents to search for vulnerable flows~\cite{repoaudit2025,nie2025vulnllm,yildiz2025jitvul}.
These formulations are usually organized around general vulnerability classes or local source-sink reasoning in the target, rather than preserving the trigger semantics of a disclosed reference case.
AI infra widens this gap because wrappers, indirection layers, and naming variation separate functional similarity from local syntactic similarity.
Therefore, \sysname faces three challenges.

\bheading{C1: How to model the semantics of the reference vulnerability and the reference and target repositories?}
Because variants may preserve a trigger mechanism while moving across modules and implementations, the system needs semantics that capture functional modules and security-relevant components while abstracting away from irrelevant details.
Previous recurring-vulnerability detectors often rely on code-reuse signatures~\cite{kim2017vuddy,woo2022movery,woo2023v1scan}, taint signatures~\cite{feng2024fire}, patch-derived signatures~\cite{li2017securitypatches,woo2022movery,huang2024vmud}, or semantic-equivalence signatures~\cite{huang2024vmud,cao2025recurring}, so their matching assumptions may not capture the specific trigger condition and semantics of a disclosed vulnerability.
Direct use of LLMs for semantic modeling of project-level code and vulnerabilities can also fail: recent project-scale studies report shallow interprocedural reasoning and misidentified source-sink pairs~\cite{li2026projscale}, and long-context models can underuse critical context buried in the prompt~\cite{liu2023lostmiddle}.

\textit{Solution:}
To address this challenge, we design a semantic modeling agent.
It uses LLM-based code understanding to construct functional modules and static analysis to model module-level call graphs.
On the repository side, this approach identifies relevant user roles and application scenarios that narrow the search space (\eg, training framework).
On the vulnerability side, it infers deployment and input boundary assumptions from the reference case (\eg, loading from an untrusted source) and maps the vulnerability to a guiding semantics that captures the trigger conditions and propagation patterns.

\bheading{C2: How should candidate inspection be localized under a bounded model context and inspection memory?}
LLM agents can interleave reasoning with tool use~\cite{yao2023react}, which makes them attractive for repository-level auditing. However, long inspections accumulate code snippets, search results, and failed hypotheses faster than the active context can retain them.
Existing repository auditors encounter similar limitations: context size, hallucination risk, and token cost must be managed by memory and on-demand exploration~\cite{repoaudit2025}.
Moreover, generic context-management techniques, such as virtual-context memory~\cite{packer2023memgpt}, reflective episodic memory~\cite{shinn2023reflexion}, and prompt compression~\cite{jiang2023llmlingua}, can preserve compact histories across long interactions.
These techniques are insufficient for reference-driven variant detection.
First, a generic summary of the inspection history may retain the agent's conclusion while dropping failed hypotheses about sources and sinks that are needed for auditability.
Second, a larger retained context still does not determine what to inspect next. Models can underuse critical context buried in long prompts~\cite{liu2023lostmiddle}, while variant inspection must spend budget according to the disclosed trigger semantics.

\textit{Solution:}
To address this challenge, we design a localization-aware state management scheme that keeps the agent's search aligned with the reference functionality under bounded context.
The key idea is to isolate critical history (\eg, searched modules and hypotheses) from the session context as cross-session inspection memory so that each session restart or context compression can resume without losing inspection-required information.
The inspection agent externalizes compact audit state, such as completed files, so later turns can avoid replaying broad repository searches.

\bheading{C3: How should final findings remain auditable given potential LLM hallucinations?}
LLM hallucination is a known failure mode in neural text generation~\cite{ji2023hallucination}, and it is especially costly in vulnerability reporting: an invented source location or exploit precondition can turn a benign path into a misleading disclosure.
Because manual verification is time-consuming, a natural baseline is to introduce an LLM judge to reason over the reported candidates.
However, an LLM judge may be biased by the candidate report itself, and hallucinated claims can be hard to distinguish without verification against the target repository.

\textit{Solution:}
To address this challenge, we implement a dynamic verification loop for candidate findings through automated PoC generation with a coding agent.
The verifier reconstructs the path from attacker-controlled input to the sensitive operation and checks whether the missing guard and preconditions preserve the reference semantics.
When execution is feasible, a sandboxed PoC records whether the exploit reaches the candidate's code path.
Otherwise, unresolved conditions remain explicit and keep the conclusion conservative.
The final report contains static code facts and verification logs needed to reproduce the finding.

\subsection{Problem Statement}
\label{sec:problem}
We consider repository code and use \emph{codebase} and \emph{repository} interchangeably throughout the paper.
We treat two commits of the same project as distinct repository revisions when their relevant semantics differ.
We next define repository semantics, reference vulnerability semantics, and vulnerability variants.

\begin{definition}[Repository semantics]
\label{def:repository-semantics}
We model the semantics of a repository $S$ as:
\begin{equation}
    \Sigma_R(S)=\langle \rho(S), \mathcal{M}(S), \phi(\mathcal{M}(S)), G_{\mathcal{M}}(S)\rangle.
\end{equation}
Here $\rho(S)$ denotes a compact summary of the repository for comparison across projects.
$\mathcal{M}(S)$ denotes functional modules based on a role taxonomy for AI infra, $\mathcal{V}_{\mathrm{role}}$. Each module $m$ has a descriptor $\phi(m)$ summarizing its assigned role and supporting code facts.
The graph $G_{\mathcal{M}}(S)$ records dependencies across modules.
\end{definition}

This abstraction follows how human auditors reason about a codebase: the repository summary supports comparison across projects, while module functionality and calling structure support bounded inspection.

\begin{definition}[Reference vulnerability semantics]
\label{def:vulnerability-semantics}
Let \refvul be a reference vulnerability discovered in a reference codebase \refrepo, described by a witness call chain \witness{\refvul} from source to sink.
We abstract \refvul into \emph{vulnerability semantics} with a fixed schema:
\begin{equation}
    \refvulnsem=\langle \psi(\witness{\refvul}), \mathcal{A}(\refrepo, \witness{\refvul})\rangle,
\end{equation}
where \(\psi\) encodes semantic features of the witness path \(\witness{\refvul}\), and \(\mathcal{A}(\refrepo, \witness{\refvul})\) represents the affected modules along \witness{\refvul}.
\end{definition}

\begin{definition}[Vulnerability variant]
\label{def:vulnerability-variant}
Given a target codebase \targetrepo, a \emph{variant} \varvul of the reference vulnerability \refvul is a vulnerability in \targetrepo whose source and sink types are semantically aligned with \refvul and whose witness path \witness{\varvul} preserves the trigger semantics of \refvul while differing in implementation details such as API migration.
\end{definition}

Our task can be formalized as follows: given a reference vulnerability \refvul with semantics \refvulnsem in a reference codebase \refrepo, detect variants \varvul in a target repository \targetrepo such that their vulnerability semantics are aligned with the reference, \(\refvulnsem\sim\varvulnsem\), and their repository semantics are similar, \ie, \(\targetreposem\sim\refreposem\).

\begin{figure*}[t]
    \centering
    \includegraphics[width=\textwidth]{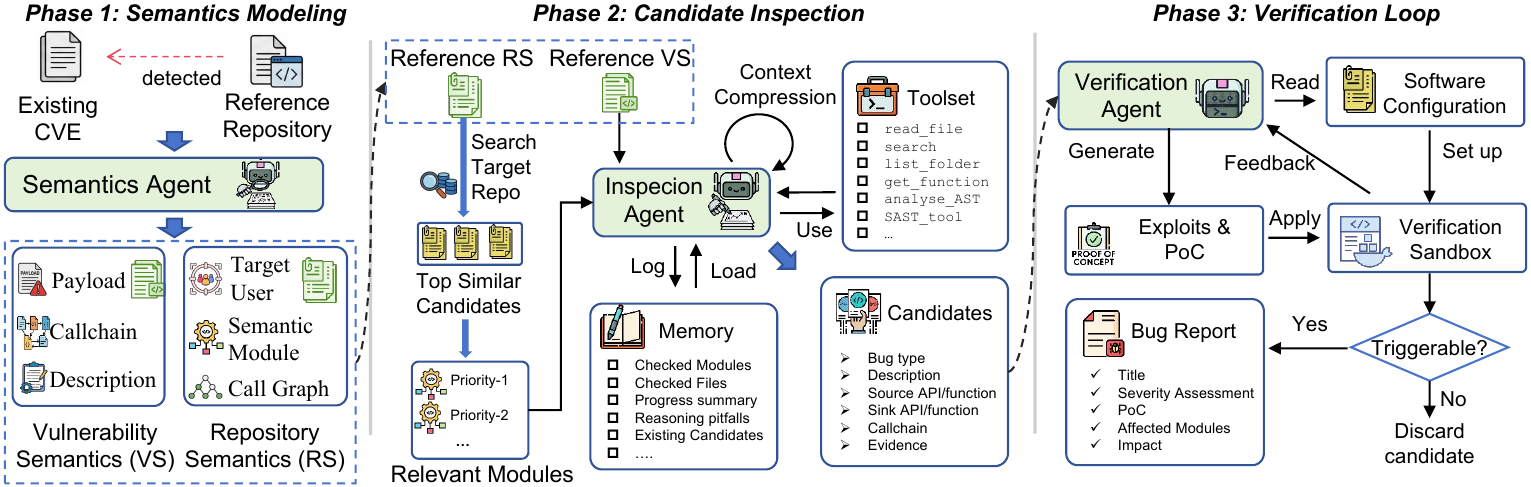}
    \caption{Overview of \sysname. The semantic agent derives vulnerability and repository semantics. Then, the inspection agent uses these semantics to prioritize target inspection under memory and context budgets. Finally, the verification agent confirms candidates with static code facts and sandbox validation before reporting findings.}
    \label{fig:method-architecture}
\end{figure*}

\section{\sysname Design}
\label{sec:method}

In this section, we present \sysname, a reference-driven auditing harness for vulnerability variants in AI infra, implemented as a controlled orchestration layer around LLM agents.

\subsection{Overview}
The harness coordinates three specialized agents: a semantic modeling agent, an inspection agent, and a verification agent.
For clarity, we present these agents in a workflow with three phases (\Cref{fig:method-architecture}), but the actual implementation can be intertwined and iterative.
Algorithm~\ref{alg:workflow} summarizes this workflow.

Given a reference vulnerability \refvul and a target codebase \targetrepo at commit $c$, \sysname first models the vulnerability semantics \refvulnsem and the repository semantics \reposem{\targetrepo,c} (Phase 1).
\(\refvulnsem\) provides downstream inspection context, such as the trigger condition and affected modules, while \(\reposem{\targetrepo,c}\) combines a repository summary $\rho(\targetrepo,c)$ for target selection across repositories with module structure for downstream localization.

In Phase 2, with the prepared vulnerability and repository semantics, \sysname selects the most similar targets based on the repository semantics to avoid unnecessary inspection, and then inspects the selected targets in priority order.
The inspection agent, provided with a set of tools that human experts would use for code inspection, iteratively searches for modules relevant to the reference vulnerability based on its semantics.
The inspection state is maintained in a local memory $\mathcal{Z}$ so that coverage, candidate state, and stopping conditions remain explicit when the live context is compacted across bounded iterations.
The inspection stops according to the configured iteration budget.

Finally, in Phase 3, the verification agent evaluates each candidate against the target repository under the assumptions required by the reference vulnerability.
To enable dynamic verification, the verification agent runs inside a coding agent that sets up a sandbox environment and generates a PoC to confirm whether the candidate can reach the same vulnerable path as the reference case.
The verifier then generates a conclusion and verification evidence (\eg, dynamic verification logs).

\subsection{Semantics Modeling}
\label{sec:semantics-modeling}

Agents perform poorly when the inspection context is underspecified.
Although advanced agents can summarize a codebase automatically, they can miss critical components (\eg, module call graphs) that support downstream variant detection.
We extract semantics for the reference vulnerability and target repository before inspection, using them as the map for downstream localization.

\bheading{Repository Semantics.}
Following Definition~\ref{def:repository-semantics}, the semantics of a repository $S$ include a repository semantic summary $\rho(S)$, modules $\mathcal{M}(S)$, functionality descriptors $\phi(m)$ for each module $m$, and a module call graph $G_{\mathcal{M}}(S)$.

Directly prompting an LLM over every file requires many tokens and is unstable across repositories.
Furthermore, without a unified category scheme, the generated module names can also differ across projects, which hinders repository comparison.
We define a role taxonomy with two levels of categories for common AI infra functionality.
This taxonomy constrains both module assignment and similarity assessment (see \Cref{tab:module-role-summary} in Appendix~\ref{appendix-subsec:module-vocabulary} for details).

The semantics agent assigns each file to one or more module roles using this taxonomy.
It first relies on path names, package structure, and local context, then reads source snippets only when the assignment is ambiguous.
For example, post-training functionality can span modules including optimizer configuration, adapter loading, checkpointing, and model loading paths.
These cases require source evidence rather than directory names alone.
During assignment, the agent stores concise descriptive features that help the downstream inspection agent decide whether a module should receive inspection budget for variants.
This process produces the modules $\mathcal{M}(S)$ and the functionality descriptors \(\{\phi(m)\}_{m\in\mathcal{M}(S)}\), each of which includes:
\begin{equation}
\phi(m)=\left\langle
\ell(m),\ \mathrm{files}(m),\ \mathrm{funcs}(m),\ \mathrm{deps}(m)
\right\rangle, 
\end{equation}
where \(\mathrm{files}(m)\), \(\mathrm{funcs}(m)\), and \(\mathrm{deps}(m)\) denote the relevant files, important functions and classes, and dependencies, respectively.

To build the dependency graph across modules $G_{\mathcal{M}}(S)$, we use an existing \ac{sast} framework (\eg, CodeQL~\cite{codeql_docs}) to extract call relations and then project them onto the assigned module descriptors \(\{\phi(m)\}_{m\in\mathcal{M}(S)}\).
The module call relations define how far inspection may expand from the matched functionality.
In addition, we extract a compact repository summary $\rho(S)$ through targeted semantic abstraction of the repository description, application scenario, target user, and key external dependencies.
This summary is used to support candidate target selection by similarity, whereas the downstream inspection agent localizes variants through $\mathcal{M}(S)$, \(\phi(\mathcal{M}(S))\), and \(G_{\mathcal{M}}(S)\).
We elaborate these repository and module similarity measures in the evaluation (\Cref{sec:evaluation}).

\bheading{Vulnerability Semantics.}
To enable transfer across projects and modules, vulnerability semantics must be narrower than a CWE class but more portable than signatures tied to one project.
They must also make explicit the assumptions under which the reference vulnerability is exploitable.
Given the call chain, payload, and affected codebase of the reference vulnerability \refvul, the agent generates \(\psi(\witness{\refvul})\) from source to sink, including the trigger condition, propagation constraints, exploitable scenario, missing guard, and affected trust boundary.
\(\psi(\witness{\refvul})\) is then used by the inspection agent to identify candidate variants and check reachability.
During this process, the agent reads relevant code snippets from \refrepo to recover affected modules \(\mathcal{A}(\refrepo, \witness{\refvul})\).
These affected modules, together with target repository semantics, feed the priority-guided inspection procedure.

\subsection{Semantics-aware Inspection}

\label{sec:prioritization}
\label{sec:agent}
Before inspection, \sysname filters and ranks candidate repositories by similarity between their semantics and the reference repository semantics, retaining targets \((\targetrepo,c)\) for which \(\reposem{\targetrepo,c}\sim\refreposem\).
In practice, we compute this similarity using information derived from both the repository summary $\rho(\cdot)$ and the module structure in \(\mathcal{M}(\cdot)\), \(\phi(\mathcal{M}(\cdot))\), and \(G_{\mathcal{M}}(\cdot)\).
For each target revision \((\targetrepo,c)\), the inspection agent uses module priority to order search and inspection memory to preserve progress.

\bheading{Priority Scheme.}
The priority scheme assumes that vulnerability variants are most likely to appear in target modules that implement functionality similar to the modules affected in the reference.
It uses three priority levels over modules.
Let $\mathcal{A}(\refrepo, \witness{\refvul})$ be the affected modules derived from the vulnerability semantics and $\mathcal{M}(\targetrepo,c)$ be the target modules from the repository semantics.
Priority 1 includes modules that implement functionality affected in the reference, including modules matching $\mathcal{A}(\refrepo, \witness{\refvul})$.
It may also promote a target module to priority 1 when the embedding similarity between its descriptor text and an affected module name exceeds a threshold $\tau_\mathcal{M}$:
\[
\begin{aligned}
    \mathcal{P}_1 &= \mathcal{P}_{name}\cup\mathcal{P}_{emb}, \text{where} \\
\mathcal{P}_{name} &= \{m\in\mathcal{M}(\targetrepo,c)\mid
    m\in\mathcal{A}(\refrepo, \witness{\refvul})\},\\
\mathcal{P}_{emb} &= \{m\in\mathcal{M}(\targetrepo,c)\mid \max_{a\in\mathcal{A}(\refrepo, \witness{\refvul})}\mathrm{sim}_{\mathcal{M}}(\phi(m),a)\ge\tau_\mathcal{M}\}.\\
\end{aligned}
\]
A module belongs to priority 2 when it directly calls a priority 1 module or is directly called by one:
\[
\mathcal{P}_2=\{m\notin\mathcal{P}_1\mid
\mathrm{calls}^{in}(m)\cap\mathcal{P}_1\ne\emptyset
\vee
\mathrm{calls}^{out}(m)\cap\mathcal{P}_1\ne\emptyset\}.
\]
Direct caller and callee expansion captures wrapper and orchestration drift while keeping the search bounded.
All remaining modules become priority 3.
The agent starts from files of $\mathcal{P}_1$ for potential sinks and sources, then $\mathcal{P}_2$, then priority 3 modules.

\bheading{Memory Management.}
To keep inspection state explicit and reuse history from previous inspections, \sysname maintains two memory layers.
The first is a local inspection memory $\mathcal{Z}$ including the status of each file within the current inspection scope, assignments from files to modules, module priorities, completion reasons, potential candidates, and the critical scope stopping boundary for $(\targetrepo,c)$.
The second is a shared public memory that stores compact observations distilled from previous inspection runs over the same target repositories, such as lightweight data flow summaries, that can avoid repeated reasoning or tool usage.
This shared layer can help later inspections avoid recomputation and redundant broad exploration.

\bheading{Iterative Inspection.}
The inspection is organized as a sequence of bounded agent iterations rather than a fixed single pass over pending files.
At iteration $t$, the agent constructs the context from $\refvulnsem$, the priority tiers, and the current state in $\mathcal{Z}$, including the progress summary, the remaining files in the configured critical scope, previously scanned files, and already recorded candidate reports.
The inspection agent then executes one bounded tool call turn over the target codebase and updates $\mathcal{Z}$ explicitly through candidate reporting and file completion actions.
A turn ends when the inspection agent emits no further tool calls, explicitly indicates completion, or is stopped by the context or turn budget.
The outer loop proceeds until the configured iteration bound and critical scope stopping policy are satisfied.
To keep enduring inspections within the model context while preserving search continuity, \sysname applies compaction before each query when the predicted request budget approaches the context window.
Each completed iteration is compressed into a compact reasoning summary that retains only inspection-relevant records, such as failures and reusable shared memory hits, while $\mathcal{Z}$ remains the source of truth for coverage and candidate state.

\begin{algorithm}[t]
    \DontPrintSemicolon
    \KwIn{Reference vulnerability \refvul with reference codebase \refrepo and target revision $(\targetrepo,c)$.}
    \KwOut{Findings verified against the target repository $\mathcal{F}$.}
    $\refreposem \leftarrow$ \textsc{ExtractRepositorySemantics}$(\refrepo)$\;
    $\refvulnsem \leftarrow$ \textsc{ExtractVulnerabilitySemantics}$(\refvul,\refreposem)$\;
    $\reposem{\targetrepo,c} \leftarrow$ \textsc{ExtractRepositorySemantics}$(\targetrepo,c)$\;
    $(\mathcal{P}_1,\mathcal{P}_2,\mathcal{P}_3) \leftarrow$ \textsc{PrioritizeModules}$(\reposem{\targetrepo,c},\refvulnsem)$\;
    $\mathcal{Z} \leftarrow$ \textsc{InspectionMemory}$(\reposem{\targetrepo,c},\refvulnsem,\mathcal{P}_1,\mathcal{P}_2,\mathcal{P}_3)$\;
    $\mathcal{C} \leftarrow \emptyset, i\leftarrow 0$\;
    \While{i < maximum iterations}{
        $i \leftarrow i + 1$\;
        Build context from $\refvulnsem$, $(\mathcal{P}_1,\mathcal{P}_2,\mathcal{P}_3)$, and $\mathcal{Z}$\;
        Run one bounded inspection turn over $(\targetrepo,c)$\;
        Update candidates $\mathcal{C}\leftarrow\mathcal{C}\cup\{\langle \mathrm{loc},\pi,\mathcal{E}_{static}\rangle\}$ and inspection memory $\mathcal{Z}$\;
    }
    $\mathcal{F}\leftarrow$ \textsc{VerifyCandidates}$(\mathcal{C},\refvulnsem,\targetrepo,c)$\;
    \Return $\mathcal{F}$\;
    \caption{Reference-driven variant inspection.}
    \label{alg:workflow}
\end{algorithm}

\subsection{Exploitability Verification}
\label{sec:verification}
The inspection agent can report a false finding due to hallucination (\eg, a nonexistent path), so the verification agent checks whether each candidate preserves the reference pattern under code facts from the target repository.
In our prototype, this stage is implemented as a reusable verification skill~\cite{agentskills2026} supported by the coding agent in \Cref{sec:evaluation}.
The skill first performs static claim checking under the assumptions recovered from the reference vulnerability and returns one of four conclusions:
\begin{packeditemize}
    \item \textit{Exploitable}: the target code contains a weakness that follows the reference pattern and supports a concrete attack scenario.
    \item \textit{Conditionally exploitable}: the weakness that follows the reference pattern is supported only under explicit input preconditions.
    \item \textit{Library risk}: a risky library dependency exists, but the current target does not expose an exploitable path.
    \item \textit{Non-exploitable}: the candidate lacks attacker-controlled source, reachability, semantic alignment, or is blocked by effective protections.
\end{packeditemize}
When the initial conclusion indicates exploitability and the target can be safely configured, the verifier generates and tests a minimal PoC in an isolated container, with multiple retries to resolve potential errors (\eg, missing dependencies).

\section{Evaluation}
\label{sec:evaluation}
In this section, we evaluate \sysname on popular real-world AI infra and investigate the following research questions:
\begin{packeditemize}
    \item \textbf{RQ1:} How accurately do the repository and vulnerability semantics align with the human labeling?
    \item \textbf{RQ2:} How does \sysname compare with baseline methods in terms of detection performance and efficiency?
    \item \textbf{RQ3:} How well does \sysname detect vulnerability variants in target repositories, and what are the main sources of false positives and false negatives?
    \item \textbf{RQ4:} How does \sysname perform on agent projects with multiple similar modules?
\end{packeditemize}

\subsection{Setup}

\bheading{LLM \& Implementation.}
Throughout the evaluation, we adopt the open-source \deepseek-V3.2~\cite{deepseekai2025deepseekv32} to ensure reproducibility.
We run local inference with \texttt{vllm} 0.18.0~\cite{vllm_docs} with decoding temperature 0.1 and top-p 0.9.
We use the official \deepseek API as a fallback when local inference fails.

We implement \sysname as a Python prototype with 29K lines.
It integrates CodeQL 2.23.7~\cite{codeql_docs} for call graph extraction and agentic SAST queries, and Claude Code (CC) 2.1.86~\cite{claude_code_docs}, powered by \deepseek-V3.2, as the coding agent that executes reusable verification skills in a controlled workspace (\ie, in the verification stage).
To select candidate repositories for inspection, we compute an overall similarity score as the equally weighted average of five components: semantic description similarity, application similarity, user similarity, module Jaccard overlap, and dependency Jaccard overlap.
For the first three components, we compare the corresponding repository profile texts with a local embedding model.
For the latter two, we compare the module sets and dependency sets derived from the repository semantics.

For each reference vulnerability, we rank all profiled target revisions by this score, keep revisions whose overall similarity is at least 0.5, and, when fewer than three revisions satisfy the threshold, supplement the scan set to the top 5 most similar revisions.
If $\refrepo$ and $\targetrepo$ differ only in commit revision, we include one additional target for inspection.
For module promotion guided by priority within each selected target, we use the default module embedding similarity threshold \(\tau_\mathcal{M}=0.8\) throughout evaluation.
We set the maximum inspection iteration to 3 per repository.
For each target repository, we extract repository semantics, build the \codeql database, run the inspection procedure, and validate all reported findings using the validation protocol.

\bheading{Reference Vulnerabilities \& Repository Candidates.}
\Cref{tab:selected-repos-latest-profile-cost-k} shows the target repository benchmark, and \Cref{tab:reference-vuln-statistics} lists the eight public reference vulnerabilities used in the evaluation.
We select the reference vulnerabilities and target repositories from the dataset in \Cref{sec:landscape} using three criteria: 1) the repository has high stars or active maintenance, 2) the disclosed data provides the call chain, affected version, and payload, and 3) the project has similar counterparts for variant detection.
After manual analysis, we select two domains, Agent \& RAG, and Training \& Inference, because they contain multiple functionally similar repositories and representative vulnerability families.
We focus on injection (\eg, SQL and command injection), SSRF, XXE, SSTI, and deserialization vulnerabilities because they are common in the dataset and can be evaluated in a sandboxed setting.

\begin{table}[t]
  \centering
  \caption{Statistics of the eight reference vulnerabilities used in the vulnerability profile accuracy evaluation. The call chain length counts source-to-sink entries.}
  \label{tab:reference-vuln-statistics}
  \resizebox{\linewidth}{!}{
  \begin{tabular}{lllrlrr}
  \toprule
  \textbf{Domain} & \textbf{Repository} & \textbf{CVE ID}  & \textbf{In/Out Tok. (K)} & \textbf{Chain} & \textbf{Vuln. Type}  & \textbf{Sim.} \\
  \midrule

  \multirow{3}{*}{Training}
  & NeMo & CVE-2025-23361 & 5.29/3.00 & 3 & Cmd. Inj. &  0.9377 \\
  & LlamaFactory & CVE-2025-53002 & 8.36/2.81 & 7 & Deser.  & 0.9095 \\
  & Megatron-LM & CVE-2025-23348 & 9.49/2.01 & 6 & Code Inj.  & 0.8791 \\
  \midrule
  \multirow{5}{*}{Agent}
  & ms-agent & CVE-2026-2256 & 8.46/2.43 & 3 & Cmd. Inj.  & 0.8849 \\
  & AutoGPT & CVE-2025-22603 & 5.58/2.57 & 4 & SSRF & 0.8815 \\
  & langchain-community & CVE-2025-6984 & 4.61/1.99 & 3 & XXE  & 0.9518 \\
  & llama\_index & CVE-2025-1793 & 4.31/1.70 & 2 & SQL Inj. & 0.8466 \\
  & AutoGPT & CVE-2025-1040 & 4.17/1.62 & 2 & SSTI & 0.8063 \\
  \bottomrule
\end{tabular}
  }
\end{table}

\begin{table}[t]
    \centering
    \caption{Target repositories as our evaluation benchmark.}
    \label{tab:selected-repos-latest-profile-cost-k}
    \resizebox{\linewidth}{!}{
    \begin{tabular}{llrrrrr}
    \toprule   
    \textbf{Type} & \textbf{Applications} & \textbf{Commit} & \textbf{Stars (K)} & \textbf{LoC (K)} & \textbf{Modules} & \textbf{In/Out Tok. (K)} \\
    \midrule                                 
    \multirow{10}{*}{Agent \& RAG}
    & AutoGPT & 42b9facd & 183.76 & 168.58 & 24 & 216.97/15.75 \\
    & langflow & e9d1c2fb & 147.35 & 228.05 & 32 & 472.56/32.35 \\
    & dify & 4461df1b & 139.15 & 596.61 & 32 & 693.51/49.46 \\
    & langchain & f2dab562 & 134.93 & 155.14 & 28 & 164.96/12.69 \\      
    & ragflow & fe4852cb & 78.98 & 190.00 & 22 & 206.60/14.59 \\                  
    & autogen & 13e144e5 & 57.43 & 79.92 & 23 & 85.57/6.83 \\
    & llama\_index & 74e5113c & 48.94 & 255.02 & 31 & 536.24/52.71 \\
    & mem0 & 84687fc3 & 54.04 & 58.28 & 17 & 60.76/4.31 \\
    & langchain-community & 39be54ca & 0.26 & 214.57 & 14 & 90.01/7.00 \\
    & ms-agent & 3d371dc7 & 4.20 & 13.22 & 12 & 14.99/1.41 \\
    \midrule
    \multirow{10}{*}{\shortstack{Training \&\\Inference}}
    & LlamaFactory & 767b344f & 70.34 & 28.91 & 21 & 31.40/2.48 \\
    & NeMo & b515e732 & 17.10 & 416.05 & 35 & 212.86/13.71 \\
    & Megatron-LM & e3ae3511 & 16.04 & 122.77 & 24 & 56.49/4.23 \\
    & ms-swift & 1a801723 & 13.90 & 64.92 & 20 & 35.12/2.66 \\
    & BentoML & 2d289a3f & 8.56 & 52.25 & 15 & 44.07/3.25 \\
    & IsaacLab & 90af2be2 & 7.00 & 111.72 & 18 & 173.68/12.75 \\
    & IsaacSim & 47d886f2 & 3.10 & 254.54 & 17 & 288.84/20.96 \\
    & GPT-SoVITS & 11aa78bd & 56.70 & 39.52 & 11 & 25.85/1.80 \\
    & Megatron-Bridge & ab225da7 & 0.59 & 45.64 & 14 & 36.44/2.59 \\
    & Model-Optimizer & b660d39a & 2.60 & 60.19 & 14 & 39.81/2.45 \\
    \bottomrule
  \end{tabular}
    }
  \end{table}
\bheading{Baselines.}
We compare against two baseline detectors: Claude Code, a fully agentic repository auditor, and Vulnhalla, a SAST pipeline augmented with LLMs~\cite{cyberark_vulnhalla,kosman2025vulnhalla}.
For Claude Code, the prompt includes the vulnerability type, source-to-sink call chain, target repository, and code inspection tools (\eg, CodeQL), and asks it to search for semantically similar vulnerabilities.
For Vulnhalla, we create CodeQL queries for each reference vulnerability family and use LLM reasoning to verify and filter CodeQL reports.

\bheading{Metrics.}
For semantic alignment, we compare generated vulnerability and repository semantics against human annotations.
For vulnerability variant detection, we merge candidates reported by all tools into a common candidate dataset and manually label each tool's decision.
We compare \sysname with the baselines using precision, recall, accuracy, and token usage.

\subsection{RQ1: Semantics Alignment}
We first examine whether the semantics align with human annotations and accurately represent the repositories and vulnerabilities.

\begin{figure*}[t]
    \centering
    \includegraphics[width=\textwidth]{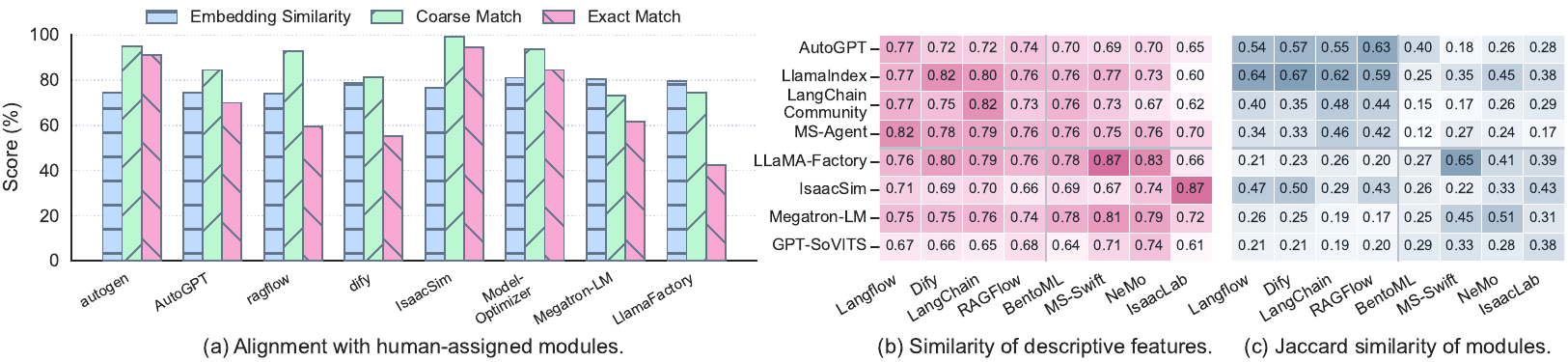}
    \caption{Similarity assessment between repository semantics. The left panel compares generated semantics with human annotations, while the right panels report descriptive and module similarity between selected target repositories.}
    \label{fig:repo-semantics-combined-figure}
\end{figure*}

\bheading{Annotation Protocol.}
We first test whether generated repository and vulnerability semantics match expert interpretation.
For vulnerability semantics, two authors summarize each reference vulnerability using the same source material, and a third author resolves the final human annotation.
For repository semantics, the same protocol is used for repository summaries in text form.
The two authors also inspect module assignments and correct the assigned roles using the predefined module role taxonomy.
The third author then resolves disagreements over the union of corrected assignments.

\bheading{Vulnerability Semantics.}
We first examine whether the extracted vulnerability semantics accurately capture the trigger semantics of the reference cases. 
As shown in \Cref{tab:reference-vuln-statistics}, the similarity scores between the descriptions annotated by humans and the generated semantics are consistently above 0.8, indicating that the generated semantics are largely aligned with human understanding after verification.
In particular, the generated semantics correctly characterize key vulnerability aspects, including vulnerability type, propagation constraints, critical sinks, and potential consequences.
The scores do not approach 1.0 mainly because the semantic descriptions generated by \deepseek are more concise than the human annotations, even when detailed contextual information is provided in the prompt.
Overall, the high semantic consistency suggests that the extracted vulnerability semantics can provide reliable guidance for subsequent variant inspection.

\bheading{Repository semantics.}
\Cref{tab:selected-repos-latest-profile-cost-k} reports the 20 target repositories used in our benchmark for vulnerability variant detection, including repository popularity, code scale at the profiled commit, the number of semantic modules extracted by our pipeline, and the reconstructed full token volume for generating the latest codebase semantics.
In total, codebase semantics generation over these repositories accounts for 3.49M input tokens and 263.98K output tokens after counting both the repository summary stage and all module analysis batches.
This is approximately 0.63 USD under the official \deepseek API pricing.
A key takeaway is that the recorded profile cost is not determined solely by repository size.
Instead, it is jointly shaped by the difficulty of identifying stable module boundaries, the diversity of subsystems within the repository, and the amount of analysis needed before the module assignments converge.
As a result, repositories with comparable or smaller code volume may still incur higher token cost if their semantic structure is harder to separate cleanly.
For instance, llama\_index has 255.02 KLoC and 31 modules, and its semantics generation consumes 536.24K input tokens. By contrast, NeMo has 416.05 KLoC and 35 modules, but consumes only 212.86K input tokens.
Likewise, langflow has 228.05 KLoC and 32 modules and consumes 472.56K input tokens, whereas IsaacSim has 254.54 KLoC and 17 modules but consumes only 288.84K input tokens.
These cases indicate that the dominant cost factor is semantic ambiguity rather than raw code volume.
We next examine whether the generated semantics are actually aligned with human perception.

\Cref{fig:repo-semantics-combined-figure} presents the alignment with modules assigned by humans on the left and the mutual similarity between the repository semantics on the right.
Overall, \Cref{fig:repo-semantics-combined-figure}(a) shows that repository semantics are aligned with human judgment at the semantic level, while exact module boundaries remain harder to reproduce.
Across all 20 repositories, the generated module descriptions reach an average BGE similarity of 77.46\% against the human annotations.
The average coarse module match is 67.31\%, and the exact module match is 51.63\%.
For the balanced subset shown in the left panel, the coarse and exact match rates increase to 86.81\% and 69.88\%, respectively.
This indicates that the generated semantics usually capture the main functional role of a component, even when the exact fine module role differs from the human annotation.

The variation mainly comes from how clearly a repository separates functional responsibilities.
For instance, IsaacSim, autogen, and Model-Optimizer have high alignment because their files are organized around stable and recognizable roles.
IsaacSim is dominated by UI workflow, dataset construction, and template components, and reaches 99.41\% coarse match and 94.45\% exact match.
Autogen similarly exposes clear agent orchestration, integration, and UI components, giving 94.95\% coarse match and 91.21\% exact match.
Model-Optimizer is also easy to annotate because most files fall into inference runtime, model export, model definition, and evaluation roles, resulting in 93.81\% coarse match and 84.54\% exact match.

The lower alignment cases show where repository semantics are still useful but human and generated boundaries diverge.
LlamaIndex has a high description similarity of 77.51\%, but only 14.29\% coarse match and 12.96\% exact match.
Its codebase contains a large number of supportive modules for agent such as vector storage and retrieval components.
Human annotations tend to group many of these files by their RAG role, while the generated semantics split or merge them according to package-level structure.
Megatron-Bridge shows a similar pattern.
Although its generated descriptions remain semantically close to the human annotations, its exact match is only 9.9\% because training configuration, model conversion, dataset building, and checkpoint handling are tightly interleaved in the repository.
These cases suggest that low agreement at the file level does not necessarily mean the semantic description is wrong.
It often reflects ambiguity in module boundaries.

In \Cref{fig:repo-semantics-combined-figure}(b) and \Cref{fig:repo-semantics-combined-figure}(c), we measure relatedness among repositories.
The four Agent \& RAG repositories occupy the upper rows and left columns, while the four Training \& Inference repositories occupy the lower rows and right columns.
The average descriptive similarity is 0.74, whereas the average module Jaccard similarity is 0.35.
Compared with the average, the blocks in the upper left and lower right have higher values.
This difference is expected because repositories can implement similar workflows using different module names and different decomposition strategies.
The high similarity pairs correspond to shared application contexts.
LLaMA-Factory and ms-swift reach 0.87 descriptive similarity and 0.65 module Jaccard similarity because both implement training and model loading workflows.
IsaacSim and IsaacLab also reach 0.87 descriptive similarity, reflecting their shared simulation context.
On the Agent and RAG side, llama\_index and dify have the highest module Jaccard similarity at 0.67, showing that both expose overlapping retrieval, integration, and workflow components.

The low similarity pairs are also meaningful.
llama\_index and IsaacLab have only 0.60 descriptive similarity, and GPT-SoVITS and IsaacLab have 0.61, because their application contexts are different even though all belong to AI infra.
Similarly, ms-agent and BentoML have a module Jaccard similarity of 0.12, indicating that an agent-tooling project and a serving framework share few module roles despite both exposing interfaces to models.
These observations support the design of \sysname, where the repository semantics preserve enough human-recognizable structure to guide localization, while the mutual similarity matrix distinguishes closely related repositories from superficially related ones.

\subsection{RQ2: Comparison with Baselines}
After checking whether generated semantics are meaningful, we evaluate whether they improve the full workflow for vulnerability variant detection.

\begin{table}[t]
      \centering
      \caption{Comparison of performance and efficiency.}
      \label{tab:compare}
      \resizebox{\linewidth}{!}{
      \begin{tabular}{lrrrrrrrr}
      \toprule
      \textbf{Method} & \textbf{TP} & \textbf{FP} & \textbf{FN} & \textbf{TN} & \textbf{Acc.} & \textbf{Prec.} & \textbf{Recall} & \textbf{In/Out Token (M)} \\
      \midrule
      Claude Code & 25 & 39 & 44 & 23 & 0.366 & 0.391 & 0.362 & 381.12/3.27 \\
      Vulnhalla~\cite{kosman2025vulnhalla} & 10 & 16 & 59 & 46 & 0.427 & 0.385 & 0.145 & 14.52/0.16 \\
      \sysname (Ours) & 24 & 7 & 45 & 55 & 0.603 & 0.774 & 0.348 & 51.11/0.58 \\
      \bottomrule
      \end{tabular}
      }
  \end{table}

\bheading{Overall Detection.}
We evaluate \sysname and baselines and gather the detected candidates.
Then, we manually label each candidate as positive (\ie, potential vulnerability) and negative (\ie, non-exploitable).
Finally, we deduplicate and unify the candidates and compare \sysname with the baselines in \Cref{tab:compare}, including candidates over the eight reference vulnerabilities.
\sysname reports 31 candidates, of which 24 are true positives and 7 are false positives, yielding the highest precision (0.774) and accuracy (0.603) among the three tools.
Its recall is lower than Claude Code because \sysname performs inspection guided by the reference and dynamic validation before retaining candidates.
This favors variants with high confidence over broad repository sweeps.

\bheading{Efficiency.}
\sysname achieves the best tradeoff between effectiveness and efficiency in this benchmark.
Claude Code is the most expensive baseline under the same workload based on reference CVEs, consuming 384.39M total tokens.
\sysname consumes 51.69M tokens, which is over 7$\times$ fewer than Claude Code, while detecting only 4\% fewer true positives.
Vulnhalla is cheaper because static scanning contributes no LLM tokens and its LLM reasoning step is smaller than agentic inspection and verification.
This lower cost comes with lower recall (0.145) and precision (0.385) than \sysname.
\sysname spends more tokens than the Vulnhalla pipeline filtered by SAST but improves both precision and recall, while using far fewer tokens than Claude Code and achieving higher accuracy.

\subsection{RQ3: Detection Performance}
The aggregate metrics show that \sysname favors precision over broad coverage.
We next examine which reference mechanisms transfer successfully and which error categories explain the remaining false positives and false negatives.
\begin{table}[t]
    \centering
    \caption{Transfer from reference to target among \sysname TP attributions. The benchmark contains 24 unique TPs; counts here are semantic attributions, and one candidate may be attributed to more than one reference CVE.}
    \label{tab:agentic-tp-transfer}
    \resizebox{\linewidth}{!}{
    \begin{tabular}{lllr}
    \toprule
    \textbf{Reference} & \textbf{Source pattern} & \textbf{Target repositories} & \textbf{\# TP} \\
    \midrule
    CVE-2025-53002 & Deserialization in adapters & BentoML, Megatron-Bridge, ms-swift & 12 \\
    CVE-2025-22603 & SSRF / URL validation bypass & autogen, langflow & 6 \\
    CVE-2025-23348 & Training asset code injection & Megatron-Bridge, NeMo, Model-Optimizer & 5 \\
    CVE-2025-1793 & Query injection & langchain-community, ragflow & 2 \\
    CVE-2025-6984 & XML / tool parsing abuse & autogen & 2 \\
    CVE-2025-1040 & Template / code injection & langflow & 3 \\
    CVE-2025-23361 & Command injection & Megatron-LM & 2 \\
    \bottomrule
    \end{tabular}
    }
\end{table}

\bheading{Transfer Across Repositories.}
\Cref{tab:agentic-tp-transfer} shows that the true positives found by \sysname are not confined to the original applications.
The strongest transfer signal comes from training and inference asset handling: the LlamaFactory deserialization reference (CVE-2025-53002) leads \sysname to unsafe model, runner, and checkpoint loading paths in BentoML, Megatron-Bridge, and ms-swift, while the NeMo and Megatron training references (CVE-2025-23361 and CVE-2025-23348) transfer to NeMo tokenizer and import utilities, Model-Optimizer conversion, TensorRT JSON serialization, and Megatron-Bridge dataset utilities, including three NeMo findings that have been assigned CVEs.
This matches the expected module semantics of AI infra: vulnerabilities in post-training and model loading code often reappear in different repositories through adapter checkpoint or dataset handling rather than through identical APIs.

\bheading{False Positives.}
Manual examination shows that the 7 false positives mainly arise when dynamic validation confirms that a sink can execute, but the final deployment boundary is not well captured by the agent and does not support a reportable vulnerability.
This occurs most often when the sink is an intended feature.
For example, \sysname treats an IsaacSim Jupyter executor that accepts code supplied by the user as suspicious, but the notebook is generally controlled by the authorized user and does not cross a new privilege boundary.
Similar reasoning applies to the ms-agent local code executor, where code or shell execution is the advertised capability rather than an unexpected privilege crossing.

The second cause involves trusted boundaries or boundaries controlled by the operator.
For example, in Model-Optimizer, the suspicious checkpoint paths are search or training assets selected by the operator rather than assets shared by an attacker under the final policy.
The third cause is source and sink overapproximation.
For instance, LangChain Cassandra, Spark SQL, and mem0 LanceDB expose query or filter construction as interfaces for callers.
In addition, some CLI paths also use safer execution forms than the source abstraction implies.
These examples explain why dynamic verification reduces hallucinated reports but cannot eliminate every false positive: the remaining ambiguity is often in deployment boundaries rather than missing code evidence.
In other words, the agent still needs stronger security policy knowledge to distinguish intended privileged functionality from an unintended attack surface.

\bheading{False Negatives.}
We also analyze the 45 missed candidates and summarize the causes and examples in \Cref{tab:addexperrortaxonomy} of Appendix~\ref{appendix:fp-and-fn-analysis}.
The main reason is that the inspection is guided by the reference vulnerability semantics, whereas the baselines rely on a looser constraint (\ie, the vulnerability type).
Furthermore, the inspection is less comprehensive than the baselines because we cap each repository at three iterations, and \deepseek sometimes stops a tool-call turn early.
GPT-SoVITS, which accounts for the most false negatives, illustrates this limitation: several unvisited files in the Web UI and audio processing modules pass user-controlled paths or text fields into shell commands.
These misses show that a repository can contain many parallel UI entries that are not all covered by a reference-guided inspection budget.
In this case, \sysname did not start the relevant file inspection before reaching the maximum iteration limit (\ie, three iterations).
When we increased the maximum number of iterations, the relevant candidates were reported after the eighth iteration, at roughly three times the cost.
Another cause is that deserialization vulnerabilities can be missed when distributed asset loaders span modules that are not captured as relevant in the repository semantics.
Overall, the false negatives are mainly coverage failures.
They indicate that \sysname is effective when the reference vulnerability points to a narrow functional context, and requires thorough references to improve coverage.

\subsection{RQ4: Case Study of OpenClaw}
The project OpenClaw has attracted broad community attention.
However, its architecture differs from the collected agent orchestration projects.
Specifically, OpenClaw has an average semantic similarity of 0.48 to the agent projects in \Cref{tab:selected-repos-latest-profile-cost-k}, lower than their mutual average of 0.53.
It also contains parallel modules with the same functionality serving different user requirements.
Therefore, it is a suitable case for testing the ability of \sysname to find variants across modules in the same project, and we focus on variants within OpenClaw (\ie, $\refrepo=\targetrepo$).

We apply \sysname to OpenClaw to examine whether one confirmed source vulnerability can guide additional target findings in the same project.
We use the GHSA-3hcm-ggvf-rch5~\cite{ghsa_3hcm_ggvf_rch5_osv} as the reference vulnerability, which is an optional \texttt{exec} approval-allowlist bypass: the command substitution or backticks hidden inside double quotes are missed by string-level allowlist analysis but interpreted by the shell during execution.
\sysname uses this execution mismatch as the reference and identifies three similar cases.
The first case, GHSA-****-****-8q4w, stays within the \texttt{exec} path: shell init-file options such as \texttt{-{}-rcfile} can inherit allowlist trust from an approved script path while loading attacker-chosen initialization first.
Similarly, \sysname transfers the same check and use drift invariant to Discord voice ingress and finds GHSA-****-****-6hch, where channel, name, and stale-role validation gaps can bypass allowlist checks over channels and members.
During examination of the inspection log, we found the agent drifts from the desired module and starts to search in irrelevant modules.
This is because our module taxonomy is not designed for OpenClaw-alike repositories.
After manually correcting this false module assignment, \sysname further identified a third case GHSA-****-****-wchg, which allows Android Canvas WebView pages from untrusted origins to invoke the \texttt{JavascriptInterface} bridge and inject instructions into the app.
Together, these findings signify that \sysname uses vulnerability semantics to preserve the source invariant, while localization guides deep inspection from an adjacent shell-wrapper variant to a cross-module authorization flaw.

\section{Related Work}
\label{sec:related-work}

\bheading{Security measurement and vulnerability evidence.}
Large-scale studies measure security patches, report reproducibility, and vulnerability lifetimes in open source software~\cite{li2017securitypatches,mu2018reproducibility,alexopoulos2022vulnerabilitylifetime}.
They show that public vulnerability evidence and patch timelines have defensive consequences, but they do not ask whether disclosed mechanisms appear as variants across functionally related AI infra repositories.
Our measurement instead characterizes functional context overlap and trigger-level similarity, then uses those findings to define an auditing task after disclosure.

\bheading{Recurring vulnerability.}
\Ac{rvd} asks whether a disclosed vulnerability or an equivalent variant reappears in another codebase.
Prior systems transfer code clones and reused components~\cite{kim2017vuddy,woo2022movery,woo2023v1scan}, vulnerable and patch signatures~\cite{li2017securitypatches,woo2022movery,huang2024vmud}, taint and firmware exploitation signatures~\cite{feng2024fire,xiao2024accurate}, semantically equivalent statements~\cite{huang2024vmud}, or black-box IoT interface signatures~\cite{yang2026iotbec}.
Cao \etal~\cite{cao2025recurring} further show that effectiveness depends strongly on recurrence characteristics and matching assumptions.
\sysname targets an orthogonal setting: the public reference case defines the semantics to preserve~\cite{zhe3,zhe4} even after significant refactoring in another AI infra project as long as the underlying vulnerability features remain similar.

\bheading{LLM-assisted vulnerability detection.}
LLM-assisted detectors use language models as analyzers of a target codebase: they audit repositories~\cite{repoaudit2025}, compare implementations with specifications~\cite{zheng2025rfcaudit}, or strengthen static analysis with taint specifications~\cite{li2025iris,zhe1,zhe2}, code property slices~\cite{lekssays2025llmxcpg}, constraints~\cite{li2025vulsolver}, and agent scaffolds~\cite{nie2025vulnllm}.
Their primary objective is to find vulnerabilities in the inspected program, not to preserve the trigger semantics of a disclosed reference case across functionally related repositories.
\sysname augments reasoning with specialized tools and harnesses for variant detection, unlocking the full reasoning potential of LLMs.

\bheading{Vulnerability reasoning evaluations.}
A separate line of work evaluates whether generic LLMs and agents can reason about vulnerabilities at code or repository scale~\cite{ullah2024secllmholmes,yildiz2025jitvul,li2026projscale}.
These studies characterize failure modes such as shallow interprocedural reasoning, incorrect guard interpretation, and high cost under project-scale context.
Long context degradation further motivates constrained localization instead of prompting over an entire repository~\cite{liu2023lostmiddle}.
Their contribution is empirical measurement of model capability, whereas \sysname turns a known vulnerability into a localized and verified cross-repository audit.

\bheading{AI software security and LLM-enabled remediation.}
LLMSmith and AgentFuzz study RCE and taint-style bugs in LLM-integrated applications and agents~\cite{liu2024llmsmith,liu2025agentfuzz}.
MirrorFuzz transfers shared bug patterns across deep learning framework APIs~\cite{ou2025mirrorfuzz}.
AppAtch uses LLMs for vulnerability patching after localization, and Vul-RAG distills vulnerability knowledge from historical vulnerable and fixed examples~\cite{nong2025appatch,du2025vulrag}.
These works either analyze AI software attack surfaces or use LLMs for post-localization repair and knowledge reuse.
They do not use a reference vulnerability to guide variant detection.

\begin{table}[t]
    \centering
    \caption{Comparison with related work. \symyes, \sympart, and \symno{} denote direct, partial, and no direct support.}
    \label{tab:related-work-positioning}
    \resizebox{\linewidth}{!}{

\begin{tabular}{lrrrrr}
\toprule
\textbf{Prior work} &
\makecell{\textbf{Uses}\\\textbf{reference}\\\textbf{semantics}} &
\makecell{\textbf{Finds}\\\textbf{related}\\\textbf{code}} &
\makecell{\textbf{Cross repo}\\\textbf{transfer}} &
\makecell{\textbf{Checks}\\\textbf{target}\\\textbf{code}} &
\makecell{\textbf{AI infra}\\\textbf{setting}} \\
\midrule

\begin{tabular}[l]{@{}l@{}}Recurring vulnerability\\ \cite{woo2023v1scan,feng2024fire,xiao2024accurate,huang2024vmud,cao2025recurring,yang2026iotbec}\end{tabular} 
& \sympart & \sympart & \symyes & \sympart & \symno \\

\rowcolor{gray!10}
\begin{tabular}[l]{@{}l@{}}LLM-assisted detection\\~\cite{repoaudit2025,zheng2025rfcaudit,li2025iris,lekssays2025llmxcpg,li2025vulsolver,nie2025vulnllm} \end{tabular}
& \symno & \sympart & \symno & \sympart & \symno \\

\begin{tabular}[l]{@{}l@{}}LLM vulnerability \\reasoning~\cite{ullah2024secllmholmes,yildiz2025jitvul,li2026projscale}\end{tabular} 
& \symno & \sympart & \symno & \sympart & \symno \\

\rowcolor{gray!10}
\begin{tabular}[l]{@{}l@{}}AI software \\vulnerability~\cite{liu2024llmsmith,liu2025agentfuzz,ou2025mirrorfuzz}\end{tabular}
& \sympart & \sympart & \sympart & \sympart & \sympart \\

\begin{tabular}[l]{@{}l@{}}LLM-based repair and\\knowledge reuse~\cite{nong2025appatch,du2025vulrag}\end{tabular}
& \sympart & \sympart & \symno & \sympart & \symno \\

\rowcolor{gray!10}
\sysname & \symyes & \symyes & \symyes & \symyes & \symyes \\

\bottomrule
\end{tabular}

}

\end{table}

\section{Conclusion and Discussion}
Through a measurement over 688 repositories and 251 related public vulnerabilities, we find that AI infra vulnerabilities often recur as variants within functionally similar modules.
This observation turns public disclosures into post-disclosure audits: whether the same trigger semantics appear in related repositories under different implementations.
We design \sysname, a reference-driven auditing harness that combines semantic modeling, localized inspection, and target-side verification.
Across our benchmark and case study, \sysname uncovers over 20 zero-day vulnerabilities, including 11 acknowledged cases and 4 assigned CVEs.

In our paper, we adopt \deepseek as the primary backbone LLM for \sysname. However, our implementation is model-agnostic and can be integrated with any LLM, as it is orthogonal to specific model capacities and relies only on the model’s reasoning and tool-use capabilities. As these capabilities, particularly context handling and reasoning, improve, the semantic modeling can become more accurate and the inspection process can be more comprehensive.

The application of \sysname is not limited to AI infra and can be extended to other domains characterized by rapidly evolving applications with multiple variants or counterparts. To enable such generalization, the module and vulnerability semantic modeling components need to become more autonomous. Currently, we rely on a domain-specific module taxonomy to extract semantics. While this approach is practically effective, it lacks generalizability when transferring to new domains without human-defined taxonomies.
As future work, we plan to leverage sub-agents to automatically construct and maintain the taxonomy from available applications, provided that sufficient diversity in the input data is ensured.

\bibliographystyle{IEEEtran}
\bibliography{ref}

@InProceedings{repoaudit2025,
  title = 	 {{R}epo{A}udit: An Autonomous {LLM}-Agent for Repository-Level Code Auditing},
  author =       {Guo, Jinyao and Wang, Chengpeng and Xu, Xiangzhe and Su, Zian and Zhang, Xiangyu},
  booktitle = 	 {Proceedings of the 42nd International Conference on Machine Learning},
  pages = 	 {21083--21100},
  year = 	 {2025},
  editor = 	 {Singh, Aarti and Fazel, Maryam and Hsu, Daniel and Lacoste-Julien, Simon and Berkenkamp, Felix and Maharaj, Tegan and Wagstaff, Kiri and Zhu, Jerry},
  volume = 	 {267},
  series = 	 {Proceedings of Machine Learning Research},
  month = 	 {13--19 Jul},
  publisher =  {PMLR},
  url = 	 {https://proceedings.mlr.press/v267/guo25n.html}
}

@inproceedings{
yao2023react,
title={ReAct: Synergizing Reasoning and Acting in Language Models},
author={Shunyu Yao and Jeffrey Zhao and Dian Yu and Nan Du and Izhak Shafran and Karthik R Narasimhan and Yuan Cao},
booktitle={The Eleventh International Conference on Learning Representations },
year={2023},
url={https://openreview.net/forum?id=WE_vluYUL-X}
}

@artical{packer2023memgpt,
  title={MemGPT: Towards LLMs as Operating Systems},
  author={Packer, Charles and Wooders, Sarah and Lin, Kevin and Fang, Vivian and Patil, Shishir G and Stoica, Ion and Gonzalez, Joseph E},
  journal={arXiv preprint arXiv:2310.08560},
  year={2023}
}

@inproceedings{shinn2023reflexion,
  title={Reflexion: language agents with verbal reinforcement learning},
  author={Noah Shinn and Federico Cassano and Ashwin Gopinath and Karthik R Narasimhan and Shunyu Yao},
  booktitle={Thirty-seventh Conference on Neural Information Processing Systems},
  volume={36},
  pages={8634--8652},
  year={2023},
  address={New Orleans, LA, USA},
  publisher={Curran Associates, Inc.},
  url={https://proceedings.neurips.cc/paper_files/paper/2023/file/1b44b878bb782e6954cd888628510e90-Paper-Conference.pdf}
}

@inproceedings{jiang2023llmlingua,
    title = "{LLML}ingua: Compressing Prompts for Accelerated Inference of Large Language Models",
    author = "Jiang, Huiqiang  and
      Wu, Qianhui  and
      Lin, Chin-Yew  and
      Yang, Yuqing  and
      Qiu, Lili",
    booktitle = "Proceedings of the 2023 Conference on Empirical Methods in Natural Language Processing",
    month = dec,
    year = "2023",
    address = "Singapore",
    publisher = "Association for Computational Linguistics",
    url = "https://aclanthology.org/2023.emnlp-main.825/",
    pages = "13358--13376"
}

@article{liu2023lostmiddle,
  title={Lost in the middle: How language models use long contexts},
  author={Liu, Nelson F and Lin, Kevin and Hewitt, John and Paranjape, Ashwin and Bevilacqua, Michele and Petroni, Fabio and Liang, Percy},
  journal={Transactions of the association for computational linguistics},
  volume={12},
  pages={157--173},
  year={2024}
}

@article{ji2023hallucination,
  author = {Ji, Ziwei and Lee, Nayeon and Frieske, Rita and Yu, Tiezheng and Su, Dan and Xu, Yan and Ishii, Etsuko and Bang, Ye Jin and Madotto, Andrea and Fung, Pascale},
  title = {Survey of Hallucination in Natural Language Generation},
  year = {2023},
  issue_date = {December 2023},
  publisher = {Association for Computing Machinery},
  address = {New York, NY, USA},
  volume = {55},
  number = {12},
  issn = {0360-0300},
  url = {https://doi.org/10.1145/3571730},
  doi = {10.1145/3571730},
  journal = {ACM Comput. Surv.},
  month = mar,
  articleno = {248},
  numpages = {38},
}

@inproceedings {liu2025agentfuzz,
author = {Fengyu Liu and Yuan Zhang and Jiaqi Luo and Jiarun Dai and Tian Chen and Letian Yuan and Zhengmin Yu and Youkun Shi and Ke Li and Chengyuan Zhou and Hao Chen and Min Yang},
title = {Make Agent Defeat Agent: Automatic Detection of {Taint-Style} Vulnerabilities in {LLM-based} Agents},
booktitle = {34th USENIX Security Symposium (USENIX Security 25)},
year = {2025},
isbn = {978-1-939133-52-6},
address = {Seattle, WA},
pages = {3767--3786},
url = {https://www.usenix.org/conference/usenixsecurity25/presentation/liu-fengyu},
publisher = {USENIX Association},
month = aug
}

@inproceedings {nong2025appatch,
author = {Yu Nong and Haoran Yang and Long Cheng and Hongxin Hu and Haipeng Cai},
title = {{APPATCH}: Automated Adaptive Prompting Large Language Models for {Real-World} Software Vulnerability Patching},
booktitle = {34th USENIX Security Symposium (USENIX Security 25)},
year = {2025},
isbn = {978-1-939133-52-6},
address = {Seattle, WA},
pages = {4481--4500},
url = {https://www.usenix.org/conference/usenixsecurity25/presentation/nong},
publisher = {USENIX Association},
month = aug
}

@inproceedings{liu2024llmsmith,
author = {Liu, Tong and Deng, Zizhuang and Meng, Guozhu and Li, Yuekang and Chen, Kai},
title = {Demystifying RCE Vulnerabilities in LLM-Integrated Apps},
year = {2024},
isbn = {9798400706363},
publisher = {Association for Computing Machinery},
address = {New York, NY, USA},
url = {https://doi.org/10.1145/3658644.3690338},
doi = {10.1145/3658644.3690338},
booktitle = {Proceedings of the 2024 on ACM SIGSAC Conference on Computer and Communications Security},
pages = {1716–1730},
numpages = {15},
location = {Salt Lake City, UT, USA},
series = {CCS '24}
}

@article{zheng2025rfcaudit,
  title={RFCAudit: An LLM Agent for Functional Bug Detection in Network Protocols},
  author={Zheng, Mingwei and Wang, Chengpeng and Liu, Xuwei and Guo, Jinyao and Feng, Shiwei and Zhang, Xiangyu},
  journal={arXiv preprint arXiv:2506.00714},
  year={2025}
}

@article{li2026projscale,
  title={LLM-based Vulnerability Detection at Project Scale: An Empirical Study},
  author={Li, Fengjie and Jiang, Jiajun and Chen, Dongchi and Xiong, Yingfei},
  journal={arXiv preprint arXiv:2601.19239},
  year={2026}
}

@INPROCEEDINGS{ullah2024secllmholmes,
  author={Ullah, Saad and Han, Mingji and Pujar, Saurabh and Pearce, Hammond and Coskun, Ayse and Stringhini, Gianluca},
  booktitle={2024 IEEE Symposium on Security and Privacy (SP)}, 
  title={LLMs Cannot Reliably Identify and Reason About Security Vulnerabilities (Yet?): A Comprehensive Evaluation, Framework, and Benchmarks}, 
  year={2024},
  volume={},
  number={},
  pages={862-880},
  doi={10.1109/SP54263.2024.00210}
}

@inproceedings{yildiz2025jitvul,
    title = "Benchmarking {LLM}s and {LLM}-based Agents in Practical Vulnerability Detection for Code Repositories",
    author = "Yildiz, Alperen  and
      Teo, Sin G  and
      Lou, Yiling  and
      Feng, Yebo  and
      Wang, Chong  and
      Divakaran, Dinil Mon",
    booktitle = "Proceedings of the 63rd Annual Meeting of the Association for Computational Linguistics (Volume 1: Long Papers)",
    month = jul,
    year = "2025",
    address = "Vienna, Austria",
    publisher = "Association for Computational Linguistics",
    url = "https://aclanthology.org/2025.acl-long.1490/",
    doi = "10.18653/v1/2025.acl-long.1490",
    pages = "30848--30865",
    ISBN = "979-8-89176-251-0",
}

@article{nie2025vulnllm,
  title={VulnLLM-R: Specialized Reasoning LLM with Agent Scaffold for Vulnerability Detection},
  author={Nie, Yuzhou and Li, Hongwei and Guo, Chengquan and Jiang, Ruizhe and Wang, Zhun and Li, Bo and Song, Dawn and Guo, Wenbo},
  journal={arXiv preprint arXiv:2512.07533},
  year={2025}
}

@inproceedings{
li2025iris,
title={{IRIS}: {LLM}-Assisted Static Analysis for Detecting Security Vulnerabilities},
author={Ziyang Li and Saikat Dutta and Mayur Naik},
booktitle={The Thirteenth International Conference on Learning Representations},
year={2025},
url={https://openreview.net/forum?id=9LdJDU7E91}
}

@inproceedings{lekssays2025llmxcpg,
  title={$\{$LLMxCPG$\}$:$\{$Context-Aware$\}$ vulnerability detection through code property $\{$Graph-Guided$\}$ large language models},
  author={Lekssays, Ahmed and Mouhcine, Hamza and Tran, Khang and Yu, Ting and Khalil, Issa},
  booktitle={34th USENIX Security Symposium (USENIX Security 25)},
  pages={489--507},
  year={2025}
}

@article{li2025vulsolver,
  title={VULSOLVER: Vulnerability Detection via LLM-Driven Constraint Solving},
  author={Li, Xiang and Su, Yueci and Liu, Jiahao and Lin, Zhiwei and Hou, Yuebing and Gao, Peiming and Zhang, Yuanchao},
  journal={arXiv preprint arXiv:2509.00882},
  year={2025}
}

@article{ou2025mirrorfuzz,
  title={MirrorFuzz: Leveraging LLM and Shared Bugs for Deep Learning Framework APIs Fuzzing},
  author={Ou, Shiwen and Li, Yuwei and Yu, Lu and Wei, Chengkun and Wen, Tingke and Chen, Qiangpu and Chen, Yu and Tang, Haizhi and Pan, Zulie},
  journal={IEEE Transactions on Software Engineering},
  year={2025},
  publisher={IEEE}
}

@misc{cve2025_1793_nvd,
  title        = {NVD - CVE-2025-1793},
  author       = {{National Vulnerability Database}},
  howpublished = {National Vulnerability Database},
  year         = {2025},
  url          = {https://nvd.nist.gov/vuln/detail/CVE-2025-1793},
  note         = {Accessed: 2026-02-20}
}

@misc{cve2025_1750_nvd,
  title        = {NVD - CVE-2025-1750},
  author       = {{National Vulnerability Database}},
  howpublished = {National Vulnerability Database},
  year         = {2025},
  url          = {https://nvd.nist.gov/vuln/detail/CVE-2025-1750},
  note         = {Accessed: 2026-02-20}
}

@misc{cve2025_53002_nvd,
  title        = {NVD - CVE-2025-53002},
  author       = {{National Vulnerability Database}},
  howpublished = {National Vulnerability Database},
  year         = {2025},
  url          = {https://nvd.nist.gov/vuln/detail/CVE-2025-53002},
  note         = {Accessed: 2026-02-20}
}

@misc{ghsa_3hcm_ggvf_rch5_osv,
  title        = {OSV: GHSA-3hcm-ggvf-rch5 (OpenClaw)},
  author       = {{Open Source Vulnerabilities (OSV)}},
  howpublished = {OSV},
  year         = {2026},
  url          = {https://osv.dev/vulnerability/GHSA-3hcm-ggvf-rch5},
}

@inproceedings{li2017securitypatches,
author = {Li, Frank and Paxson, Vern},
title = {A Large-Scale Empirical Study of Security Patches},
year = {2017},
isbn = {9781450349468},
publisher = {Association for Computing Machinery},
address = {New York, NY, USA},
url = {https://doi.org/10.1145/3133956.3134072},
doi = {10.1145/3133956.3134072},
booktitle = {Proceedings of the 2017 ACM SIGSAC Conference on Computer and Communications Security},
pages = {2201–2215},
numpages = {15},
location = {Dallas, Texas, USA},
series = {CCS '17}
}

@inproceedings{mu2018reproducibility,
  author = {Dongliang Mu and Alejandro Cuevas and Limin Yang and Hang Hu and Xinyu Xing and Bing Mao and Gang Wang},
  title = {Understanding the Reproducibility of Crowd-reported Security Vulnerabilities},
  booktitle = {27th USENIX Security Symposium (USENIX Security 18)},
  year = {2018},
  isbn = {978-1-939133-04-5},
  address = {Baltimore, MD},
  pages = {919--936},
  url = {https://www.usenix.org/conference/usenixsecurity18/presentation/mu},
  publisher = {USENIX Association},
  month = aug
}

@inproceedings{alexopoulos2022vulnerabilitylifetime,
author = {Nikolaos Alexopoulos and Manuel Brack and Jan Philipp Wagner and Tim Grube and Max M{\"u}hlh{\"a}user},
title = {How Long Do Vulnerabilities Live in the Code? A {Large-Scale} Empirical Measurement Study on {FOSS} Vulnerability Lifetimes},
booktitle = {31st USENIX Security Symposium (USENIX Security 22)},
year = {2022},
isbn = {978-1-939133-31-1},
address = {Boston, MA},
pages = {359--376},
url = {https://www.usenix.org/conference/usenixsecurity22/presentation/alexopoulos},
publisher = {USENIX Association},
month = aug
}

@inproceedings{kim2017vuddy,
  author={Kim, Seulbae and Woo, Seunghoon and Lee, Heejo and Oh, Hakjoo},
  booktitle={2017 IEEE Symposium on Security and Privacy (SP)}, 
  title={VUDDY: A Scalable Approach for Vulnerable Code Clone Discovery}, 
  year={2017},
  volume={},
  number={},
  pages={595-614},
  doi={10.1109/SP.2017.62}}

@inproceedings{woo2022movery,
  author = {Seunghoon Woo and Hyunji Hong and Eunjin Choi and Heejo Lee},
  title = {{MOVERY}: A Precise Approach for Modified Vulnerable Code Clone Discovery from Modified {Open-Source} Software Components},
  booktitle = {31st USENIX Security Symposium (USENIX Security 22)},
  year = {2022},
  isbn = {978-1-939133-31-1},
  address = {Boston, MA},
  pages = {3037--3053},
  url = {https://www.usenix.org/conference/usenixsecurity22/presentation/woo},
  publisher = {USENIX Association},
  month = aug
}

@misc{cyberark_vulnhalla,
  title        = {Vulnhalla},
  author       = {{CyberArk}},
  howpublished = {GitHub repository},
  year         = {2025},
  url          = {https://github.com/cyberark/Vulnhalla},
  note         = {Accessed: 2026-03-01}
}

@misc{kosman2025vulnhalla,
  title        = {Vulnhalla: Picking the True Vulnerabilities from a CodeQL Haystack},
  author       = {Kosman, Simcha},
  howpublished = {CyberArk Threat Research Blog},
  year         = {2025},
  month        = dec,
  url          = {https://www.cyberark.com/resources/threat-research-blog/vulnhalla-picking-the-true-vulnerabilities-from-the-codeql-haystack},
}

@article{du2025vulrag,
  title={Vul-rag: Enhancing llm-based vulnerability detection via knowledge-level rag},
  author={Du, Xueying and Zheng, Geng and Wang, Kaixin and Zou, Yi and Wang, Yujia and Deng, Wentai and Feng, Jiayi and Liu, Mingwei and Chen, Bihuan and Peng, Xin and others},
  journal={ACM Transactions on Software Engineering and Methodology},
  year={2024},
  publisher={ACM New York, NY}
}

@misc{mythos2026,
  title        = {Claude Mythos Preview},
  author       = {{Anthropic}},
  howpublished = {Anthropic Alignment Science technical blog},
  year         = {2026},
  month        = apr,
  url          = {https://red.anthropic.com/2026/mythos-preview/},
  note         = {Published: 2026-04-07. Accessed: 2026-04-08}
}

@inproceedings{xiao2024accurate,
  author = {Xiao, Haoyu and Zhang, Yuan and Shen, Minghang and Lin, Chaoyang and Zhang, Can and Liu, Shengli and Yang, Min},
  title = {Accurate and Efficient Recurring Vulnerability Detection for IoT Firmware},
  year = {2024},
  isbn = {9798400706363},
  publisher = {Association for Computing Machinery},
  address = {New York, NY, USA},
  url = {https://doi.org/10.1145/3658644.3670275},
  doi = {10.1145/3658644.3670275},
  pages = {3317–3331},
  numpages = {15},
  location = {Salt Lake City, UT, USA},
  series = {CCS '24}
}

@inproceedings{woo2023v1scan,
  author = {Seunghoon Woo and Eunjin Choi and Heejo Lee and Hakjoo Oh},
  title = {{V1SCAN}: Discovering 1-day Vulnerabilities in Reused {C/C++} Open-source Software Components Using Code Classification Techniques},
  booktitle = {32nd USENIX Security Symposium (USENIX Security 23)},
  year = {2023},
  isbn = {978-1-939133-37-3},
  address = {Anaheim, CA},
  pages = {6541--6556},
  url = {https://www.usenix.org/conference/usenixsecurity23/presentation/woo},
  publisher = {USENIX Association},
  month = aug
}

@inproceedings{feng2024fire,
  author = {Siyue Feng and Yueming Wu and Wenjie Xue and Sikui Pan and Deqing Zou and Yang Liu and Hai Jin},
  title = {{FIRE}: Combining {Multi-Stage} Filtering with Taint Analysis for Scalable Recurring Vulnerability Detection},
  booktitle = {33rd USENIX Security Symposium (USENIX Security 24)},
  year = {2024},
  isbn = {978-1-939133-44-1},
  address = {Philadelphia, PA},
  pages = {1867--1884},
  url = {https://www.usenix.org/conference/usenixsecurity24/presentation/feng-siyue},
  publisher = {USENIX Association},
  month = aug
}

@inproceedings{huang2024vmud,
author = {Huang, Kaifeng and Lu, Chenhao and Cao, Yiheng and Chen, Bihuan and Peng, Xin},
title = {VMud: Detecting Recurring Vulnerabilities with Multiple Fixing Functions via Function Selection and Semantic Equivalent Statement Matching},
year = {2024},
isbn = {9798400706363},
publisher = {Association for Computing Machinery},
address = {New York, NY, USA},
url = {https://doi.org/10.1145/3658644.3690372},
doi = {10.1145/3658644.3690372},
booktitle = {Proceedings of the 2024 on ACM SIGSAC Conference on Computer and Communications Security},
pages = {3958–3972},
numpages = {15},
location = {Salt Lake City, UT, USA},
series = {CCS '24}
}

@article{cao2025recurring,
author = {Cao, Yiheng and Wu, Susheng and Wang, Ruisi and Chen, Bihuan and Huang, Yiheng and Lu, Chenhao and Zhou, Zhuotong and Peng, Xin},
title = {Recurring Vulnerability Detection: How Far Are We?},
year = {2025},
issue_date = {July 2025},
publisher = {Association for Computing Machinery},
address = {New York, NY, USA},
volume = {2},
number = {ISSTA},
url = {https://doi.org/10.1145/3728901},
doi = {10.1145/3728901},
journal = {Proc. ACM Softw. Eng.},
month = jun,
articleno = {ISSTA026},
numpages = {23},
}

@inproceedings{yang2026iotbec,
  title={IoTBec: An Accurate and Efficient Recurring Vulnerability Detection Framework for Black Box IoT devices},
  author={Yang, Haoran and Guo, Jiaming and Yang, Shuangning and Zhao, Guoli and Liu, Qingqi and Zhang, Chi and Tan, Zhenlu and Shan, Lixiao and Zhou, Qihang and Zhou, Mengting and Tai, Jianwei and Jia, Xiaoqi},
  booktitle={Proceedings of the Network and Distributed System Security Symposium (NDSS)},
  year={2026},
  publisher={The Internet Society},
  address={San Diego, CA, USA},
  numpages={17},
  doi={10.14722/ndss.2026.240634}
}

@misc{deepseekai2025deepseekv32,
      title={DeepSeek-V3.2: Pushing the Frontier of Open Large Language Models}, 
      author={DeepSeek-AI},
      year={2025},
}

@misc{bge_embedding,
      title={C-Pack: Packaged Resources To Advance General Chinese Embedding}, 
      author={Shitao Xiao and Zheng Liu and Peitian Zhang and Niklas Muennighoff},
      year={2023},
      eprint={2309.07597},
      archivePrefix={arXiv},
      primaryClass={cs.CL}
}

@article{sheng2024hybridflow,
author = {Sheng, Guangming and Zhang, Chi and Ye, Zilingfeng and Wu, Xibin and Zhang, Wang and Zhang, Ru and Peng, Yanghua and Lin, Haibin and Wu, Chuan},
title = {HybridFlow: A Flexible and Efficient RLHF Framework},
year = {2025},
isbn = {9798400711961},
publisher = {Association for Computing Machinery},
address = {New York, NY, USA},
url = {https://doi.org/10.1145/3689031.3696075},
doi = {10.1145/3689031.3696075},
booktitle = {Proceedings of the Twentieth European Conference on Computer Systems},
pages = {1279–1297},
numpages = {19},
location = {Rotterdam, Netherlands},
series = {EuroSys '25}
}

@inproceedings{zheng-etal-2024-llamafactory,
    title = "{L}lama{F}actory: Unified Efficient Fine-Tuning of 100+ Language Models",
    author = "Zheng, Yaowei  and
      Zhang, Richong  and
      Zhang, Junhao  and
      Ye, Yanhan  and
      Luo, Zheyan",
    booktitle = "Proceedings of the 62nd Annual Meeting of the Association for Computational Linguistics (Volume 3: System Demonstrations)",
    month = aug,
    year = "2024",
    address = "Bangkok, Thailand",
    publisher = "Association for Computational Linguistics",
    url = "https://aclanthology.org/2024.acl-demos.38/",
    doi = "10.18653/v1/2024.acl-demos.38",
    pages = "400--410"
}

@misc{agentskills2026,
  title        = {Agent Skills},
  author       = {{Agent Skills}},
  year         = {2026},
  howpublished = {Official documentation},
  url          = {https://agentskills.io/home},
  note         = {Accessed: 2026-04-22}
}

@misc{nvd_general,
  title        = {National Vulnerability Database},
  author       = {{National Institute of Standards and Technology}},
  howpublished = {Official vulnerability database},
  year         = {2026},
  url          = {https://nvd.nist.gov/},
  note         = {Accessed: 2026-04-27}
}

@misc{osv_dev,
  title        = {Open Source Vulnerabilities},
  author       = {{Open Source Vulnerabilities}},
  howpublished = {Official vulnerability database},
  year         = {2026},
  url          = {https://osv.dev/},
  note         = {Accessed: 2026-04-27}
}

@misc{github_advisory_database,
  title        = {GitHub Advisory Database},
  author       = {{GitHub}},
  howpublished = {Official vulnerability advisory database},
  year         = {2026},
  url          = {https://github.com/advisories},
  note         = {Accessed: 2026-04-27}
}

@misc{huntr_platform,
  title        = {Huntr},
  author       = {{Huntr}},
  howpublished = {Vulnerability disclosure platform},
  year         = {2026},
  url          = {https://huntr.com/},
  note         = {Accessed: 2026-04-27}
}

@misc{codeql_docs,
  title        = {CodeQL Documentation},
  author       = {{GitHub}},
  howpublished = {Official documentation},
  year         = {2026},
  url          = {https://codeql.github.com/docs/},
  note         = {Accessed: 2026-04-27}
}

@misc{claude_code_docs,
  title        = {Claude Code Documentation},
  author       = {{Anthropic}},
  howpublished = {Official documentation},
  year         = {2026},
  url          = {https://code.claude.com/docs/},
  note         = {Accessed: 2026-04-27}
}

@misc{vllm_docs,
  title        = {{vLLM} Documentation},
  author       = {{vLLM Project}},
  howpublished = {Official documentation},
  year         = {2026},
  url          = {https://docs.vllm.ai/},
  note         = {Accessed: 2026-04-27}
}

@article{zhe1,
author = {Zhang, Huaien and Pei, Yu and Liang, Shuyun and Tan, Shin Hwei},
title = {Understanding and Detecting Annotation-Induced Faults of Static Analyzers},
year = {2024},
issue_date = {July 2024},
publisher = {Association for Computing Machinery},
address = {New York, NY, USA},
volume = {1},
number = {FSE},
url = {https://doi.org/10.1145/3643759},
doi = {10.1145/3643759},
journal = {Proc. ACM Softw. Eng.},
month = jul,
articleno = {33},
numpages = {23},
}

@inproceedings{zhe2,
author = {Zhang, Huaien and Pei, Yu and Liang, Shuyun and Xing, Zezhong and Tan, Shin Hwei},
title = {Characterizing and Detecting Program Representation Faults of Static Analysis Frameworks},
year = {2024},
isbn = {9798400706127},
publisher = {Association for Computing Machinery},
address = {New York, NY, USA},
url = {https://doi.org/10.1145/3650212.3680398},
doi = {10.1145/3650212.3680398},
booktitle = {Proceedings of the 33rd ACM SIGSOFT International Symposium on Software Testing and Analysis},
pages = {1772–1784},
numpages = {13},
location = {Vienna, Austria},
series = {ISSTA 2024}
}

@inproceedings{zhe3,
author = {Zhang, Huaien and Pei, Yu and Chen, Junjie and Tan, Shin Hwei},
title = {Statfier: Automated Testing of Static Analyzers via Semantic-Preserving Program Transformations},
year = {2023},
isbn = {9798400703270},
publisher = {Association for Computing Machinery},
address = {New York, NY, USA},
url = {https://doi.org/10.1145/3611643.3616272},
doi = {10.1145/3611643.3616272},
booktitle = {Proceedings of the 31st ACM Joint European Software Engineering Conference and Symposium on the Foundations of Software Engineering},
pages = {237–249},
numpages = {13},
location = {San Francisco, CA, USA},
series = {ESEC/FSE 2023}
}

@article{zhe4,
author = {Wang, Haibo and Xu, Zhuolin and Zhang, Huaien and Tsantalis, Nikolaos and Tan, Shin Hwei},
title = {Towards Understanding Refactoring Engine Bugs},
year = {2026},
issue_date = {May 2026},
publisher = {Association for Computing Machinery},
address = {New York, NY, USA},
volume = {35},
number = {5},
issn = {1049-331X},
url = {https://doi.org/10.1145/3747289},
doi = {10.1145/3747289},
journal = {ACM Trans. Softw. Eng. Methodol.},
month = apr,
articleno = {138},
numpages = {55},
}

@article{infraforagent_tmlr25,
  author       = {Alan Chan and
                  Kevin Wei and
                  Sihao Huang and
                  Nitarshan Rajkumar and
                  Elija Perrier and
                  Seth Lazar and
                  Gillian K. Hadfield and
                  Markus Anderljung},
  title        = {Infrastructure for {AI} Agents},
  journal      = {Trans. Mach. Learn. Res.},
  volume       = {2025},
  year         = {2025},
  url          = {https://openreview.net/forum?id=Ckh17xN2R2},
  bibsource    = {dblp computer science bibliography, https://dblp.org}
}

\appendices

\section{Ethical Considerations}

This study relies only on the public repositories and publicly disclosed vulnerabilities.
All validation is performed on local checkouts and isolated sandbox environments controlled by the authors.
We do not scan production services, interact with live deployments, test assets without authorization, or collect private user data.
The study does not involve human subjects or private personal data, so institutional review board review is not applicable under this scope.

The main beneficiaries are maintainers, downstream users, and defensive researchers who need to determine whether a disclosed vulnerability mechanism has variants in related AI infra repositories.
The main risk is that information about similar mechanisms or validation paths could help attackers target projects whose vulnerabilities are not yet fixed.
A secondary risk is that generated exploit code could be mishandled during analysis.
We reduce these risks by limiting the study to repository code and public disclosures, running validation locally, and treating generated PoCs as restricted materials.

\sysname reports a candidate only after verification against the target repository, so the LLM inference alone is never treated as a vulnerability claim.
For findings that remain unfixed or are still under coordination, we withhold payloads, detailed traces, logs for individual cases, and identifiers that would make exploitation or target selection easier.
We judge the remaining publication risk to be limited relative to its defensive value, because the paper reports aggregate methodology and results without exposing unresolved exploit details.

\section{Measurement Protocol}
\label{sec:appendix-training-cluster-measurement}

\subsection{GitHub Search and Curation}

The broad repository search uses the GitHub Search REST API to query public GitHub repositories that are not archived and were created on or after January 1, 2023.
Topic queries use topic seeds, and keyword queries search quoted seeds over repository names, descriptions, and README files.
The seeds cover agent-orchestration projects, LLMOps, RAG, serving, inference, observability, retrieval pipelines, workflow runtimes, and tool calling.
For this family, the topic seeds are ``ai-agent'', ``agent-framework'', and ``agent-orchestration''.
Other seeds include ``llmops'', ``rag'', ``model-serving'', ``inference-server'', ``ai-infrastructure'', and ``observability''.
Retrieval pipelines, workflow runtimes, and tool calling use keyword seeds only.
The keyword seeds are ``rag'' and ``retrieval pipeline'' for RAG, ``model serving'' and ``inference engine'' for serving and inference, ``agent orchestration'' and ``workflow runtime'' for agent orchestration, and ``llm observability'' and ``tool calling'' for observability.

Before aggregation, we keep the GitHub information used in the paper and merge repeated results by repository.
We then add direct AI infra repositories that appear in the public vulnerability disclosure dataset but were not returned by the 2023+ discovery queries, and record this provenance in the repository audit table.
Two authors independently checked whether each retained repository provides reusable AI infra functionality and assigned a primary capability family, with a third author resolving disagreements.
The exclusion rules remove tutorials and courseware, sample and template repositories, prompt collections, benchmark repositories, personal assistant or chat applications, generic UI or proxy tools without reusable AI infra functionality, model weight dumps, mirrors and translations, and showcases without substantive reusable infrastructure logic.

\subsection{Training Cluster Measurement Protocol}

\begin{table}[t]
    \centering
    \small
    \setlength{\tabcolsep}{4.2pt}
    \begin{tabular}{lcc}
        \toprule
        Feature & Stacks & Share \\
        \midrule
        Checkpoint recovery & 6 & 85.7\% \\
        FSDP or ZeRO & 6 & 85.7\% \\
        Dependencies and environments & 6 & 85.7\% \\
        Tests & 6 & 85.7\% \\
        vLLM integration & 6 & 85.7\% \\
        Quantization and export & 5 & 71.4\% \\
        Container configurations & 4 & 57.1\% \\
        Web interface & 3 & 42.9\% \\
        \bottomrule
    \end{tabular}
    \caption{Repeated training stack features. The table reports repeated code and workflow features across the same seven frameworks. Stacks (of 7) counts how many of the seven frameworks contain each feature. Cohort share is the same count divided by seven.}
    \label{tab:training-cluster-surfaces}
\end{table}

This section reports the repository collection and source tree listings used for the seven framework training cluster analysis in \Cref{fig:training-cluster-similarity,tab:training-cluster-surfaces}.
The cohort contains LLaMA-Factory, ms-swift, axolotl, TRL, verl, OpenRLHF, and unsloth.
For each project, we use the GitHub default branch as of April 2, 2026 and retrieve the recursive source tree listing for that branch.
The measurement covers source tree structure and functional indicators on the default branch.

We encode each repository with 12 binary features spanning project structure and training workflow motifs.
The structure side covers documentation and examples, tests, container and runtime configs, requirements files, \texttt{pyproject.toml}, setup files, and \texttt{src/} layout.
The motif side covers checkpoint and resume logic, Web UI interfaces, quantization and export paths, FSDP or ZeRO runtime support, and vLLM integration.
Each feature is identified from repository paths using predefined indicators.
Representative path indicators include tests, requirements files, Docker directories, checkpoints, \texttt{gradio}, and \texttt{vllm}.
Accordingly, this is a measurement over repository paths rather than full semantic parsing. It should be read as a proxy for implementation features, not proof of identical behavior.

From this binary matrix, \Cref{fig:training-cluster-similarity} reports lower triangle shared feature counts over the 12 binary features.
\Cref{fig:training-cluster-similarity} also reports upper triangle README embedding similarities computed from the same seven framework cohort.
\Cref{tab:training-cluster-surfaces} reports the motif prevalence table over the eight features used for the main text summary.
In that table, Stacks (of 7) counts how many of the seven repositories contain the feature, and Cohort share reports the same count normalized by the seven repository cohort.
Within this measurement range, counts of shared features are bounded by the same 12 binary features, while motif prevalence ranges from 0 to 7 repositories. In the current cohort, five of the eight measured features appear in 6 of 7 stacks and none appears in all seven.

\section{Supplementary Results}

\subsection{Variant within the Same Project}

\begin{figure*}[t]
    \centering
    \includegraphics[width=0.85\linewidth]{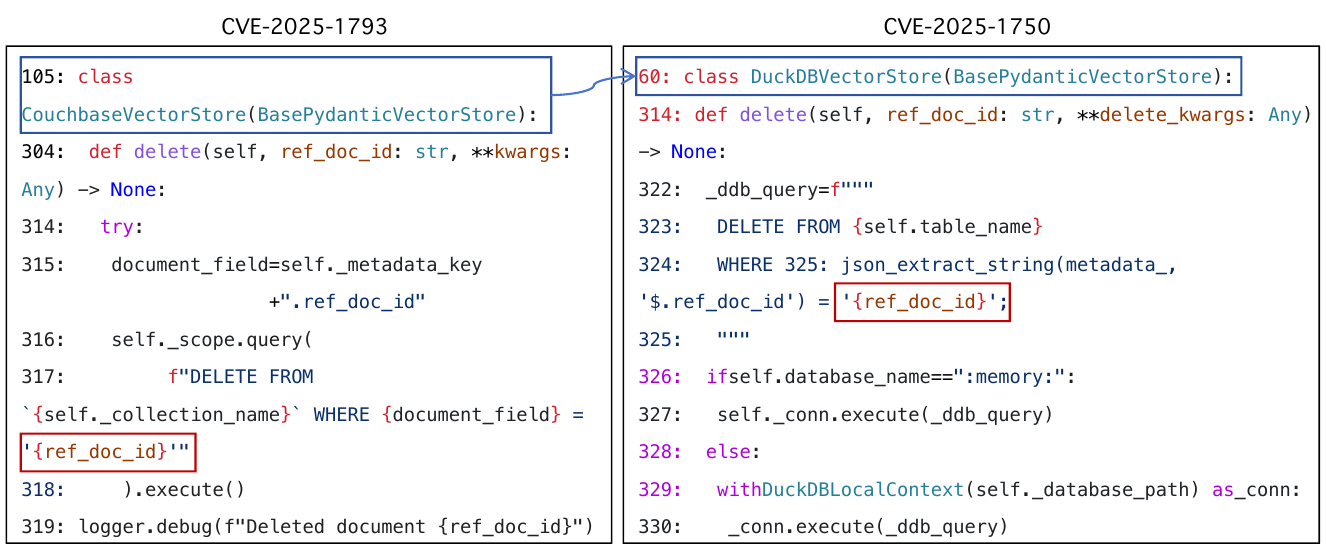}
    \caption{SQL injection variants between two modules within \texttt{llama\_index}. CVE-2025-1793 and CVE-2025-1750 (both are public records) occur in Couchbase and DuckDB vector store integrations but interpolate attacker-controlled document identifiers into SQL DELETE statements.}
    \label{fig:within-project-variant}
\end{figure*}

\Cref{fig:within-project-variant} shows variants within \texttt{llama\_index}.
CVE-2025-1793 affects a Couchbase vector store integration, while CVE-2025-1750 affects a DuckDB vector store integration~\cite{cve2025_1793_nvd,cve2025_1750_nvd}.
The implementations differ in backend API and file location, but both delete paths interpolate a document identifier controlled by an attacker into an SQL \texttt{DELETE} statement.
This case complements the main text example across repositories: variants can appear across modules inside one project as well as across related AI infra projects.

\subsection{Inspection Agent Tools}
\label{appendix-subsec:inspection-agent-tools}
\begin{table}[t]
    \centering
    \scriptsize
    \setlength{\tabcolsep}{3pt}
    \begin{tabular}{p{0.35\linewidth}p{0.59\linewidth}}
        \toprule
        \textbf{Tool} & \textbf{Description} \\
        \midrule
        \makecell[l]{\path{read_file} \\\\ \path{get_function_code}} & Inspect code with line numbers and extract function/class bodies. \\
        \midrule
        \makecell[l]{\path{search_in_file} \\\\ \path{search_in_folder}} & Locate sources, sinks, and semantic patterns via regex search. \\
        \midrule
        \makecell[l]{\path{list_files_in_folder} \\\\ \path{get_imports}} & Quickly summarize a module and its dependencies. \\
        \midrule
        \path{analyze_data_flow} & Summarize intraprocedural data propagation (parameters, assignments, calls, returns). \\
        \midrule
        \makecell[l]{\texttt{get\_related\_files} \\\\ \texttt{get\_module\_call\_} \\\\ \texttt{relationships}} & Expand analysis along static call relationships from the repository semantics. \\
        \midrule
        \makecell[l]{\path{run_codeql_query} \\\\ \path{read_codeql_results}} & Execute \codeql queries and consume SARIF findings for verification. \\
        \midrule
        \makecell[l]{\texttt{read\_shared} \\\\ \texttt{public\_memory}} & Load prior observations as localization hints for the same target scope. \\
        \midrule
        \path{report_vulnerability} & Emit a structured candidate with location, supporting code facts, and confidence for later verification. \\
        \midrule
        \makecell[l]{\path{mark_file_completed} \\\\ \path{check_file_status}} & Track progress in persistent analysis state and avoid duplicate inspection. \\
        \bottomrule
    \end{tabular}
    \caption{Codebase analysis interface used by the inspection procedure.}
    \label{tab:tools}
\end{table}
\bheading{Agentic Tools.}
\Cref{tab:tools} shows the tools used by the inspection agent.
The tool design follows the constrained interfaces exposed by the current coding agent~\cite{claude_code_docs} while keeping each action auditable.

\subsection{False Positive and False Negative}
\label{appendix:fp-and-fn-analysis}

\Cref{tab:addexperrortaxonomy} summarizes the manual judgment categories behind the FP and FN analysis in RQ3, where FP rows summarize reported candidates rejected after human review, and FN rows summarize missed variants in the benchmark.
\begin{table*}[t]
    \centering
    \caption{Summary of FP and FN categories. Examples are representative repository contexts used to explain the FP and FN analysis.}
    \label{tab:addexperrortaxonomy}
    \small
    \setlength{\tabcolsep}{3pt}
    \begin{tabularx}{\textwidth}{>{\raggedright\arraybackslash}p{0.08\textwidth}>{\raggedright\arraybackslash}p{0.22\textwidth}>{\raggedright\arraybackslash}X>{\raggedright\arraybackslash}p{0.32\textwidth}}
    \toprule
    \textbf{Outcome} & \textbf{Category} & \textbf{Review cue} & \textbf{Representative examples} \\
    \midrule
    FP & Intended privileged functionality & Advertised capability for an authorized user & \makecell[l]{IsaacSim Jupyter executor\\\texttt{ms-agent} local executor} \\
    \midrule
    FP & Trusted deployment boundary & Operator-controlled data or internal service control required & \makecell[l]{BentoML remote runner response\\Model-Optimizer search checkpoint\\Megatron-Bridge checkpoint state} \\
    \midrule
    FP & Source and sink overapproximation & Caller-facing query or filter interfaces and safer CLI execution weaken attacker-control or sink assumptions & \makecell[l]{\texttt{mem0} LanceDB filters\\LLaMA-Factory SGLang argv launch} \\
    
    \midrule
    FN & Broad UI command paths & User controlled paths appear outside the localized budget & \makecell[l]{GPT-SoVITS Web UI\\UVR audio processing paths} \\
    FN & Distributed data loaders & Checkpoint and adapter loaders span modules & \makecell[l]{GPT-SoVITS checkpoint loaders\\Transformers loaders\\Megatron-LM checkpointing} \\
    FN & URL fetch and SSRF paths & Agent tools and workflow components expose many fetch paths & \makecell[l]{AutoGPT request utility\\Dify HTTP request executor\\Langflow RSS and web search} \\
    FN & Vector-store query construction & Query syntax is assembled in adapter-specific storage layers & \makecell[l]{LangChain MyScale and SingleStoreDB\\LlamaIndex Couchbase} \\
    \bottomrule
    \end{tabularx}
\end{table*}
\subsection{Module Role Taxonomy}
\label{appendix-subsec:module-vocabulary}

\Cref{tab:module-role-summary} shows the role taxonomy used for repository semantics extraction.
We construct this taxonomy in three steps.
First, three annotators label the top three reusable functionalities in 100 crawled repositories, producing 300 coarse raw module labels.
Second, annotators group related labels and summarize disagreements within each group.
Third, annotators decide whether each coarse category is necessary and keep at most five common second-level roles for retained categories.
This procedure supports generalization across projects through coarse module categories while preserving domain-specific distinctions through second-level roles.

\begin{table*}[t]
    \centering
    \scriptsize
    \setlength{\tabcolsep}{2pt}
    \renewcommand{\arraystretch}{0.9}
    \begin{tabularx}{\textwidth}{@{}>{\raggedright\arraybackslash}p{0.22\textwidth}>{\raggedright\arraybackslash}X>{\raggedright\arraybackslash}p{0.34\textwidth}@{}}
        \toprule
        \textbf{Module Categorization} & \textbf{Definition} & \textbf{Second Level Module Role} \\
        \midrule
        Platform Systems & Build, package, configure, and orchestrate the runtime substrate for local, containerized, or distributed AI workloads. & Build Packaging \\
         & & Runtime Hardware \\
         & & Distributed Orchestration \\
        \addlinespace[1pt]
        Data Knowledge & Ingest, normalize, chunk, store, and retrieve datasets or external knowledge used by models and applications. & Ingestion Connectors \\
         & & Dataset Construction \\
         & & Preprocess Tokenization \\
         & & Storage Formats \\
         & & Knowledge Stores \\
        \addlinespace[1pt]
        Model Assets and Loading & Define model assets and loading paths, including architectures, checkpoints, tokenizers, processors, and runtime configuration. & Model Definition \\
         & & Checkpoint Formats \\
         & & Loading Configuration \\
         & & Tokenizers/Processors \\
         & & Export Interchange \\
        \addlinespace[1pt]
        Training and Optimization & Run training loops, distributed optimization, checkpointing, and experiment configuration. & Training Loop \\
         & & Distributed Training \\
         & & Optimizer Schedules \\
         & & Checkpoint/Finetuning \\
         & & Experiment Configurations \\
        \addlinespace[1pt]
        Post-Training and Alignment & Adapt pretrained models through supervised finetuning, PEFT, preference optimization, RLHF/RLAIF, or distillation style procedures. & Supervised Finetuning \\
         & & Parameter Efficient Finetuning \\
         & & Preference Learning \\
         & & RLHF/RLAIF \\
         & & Distillation/Quantization Aware Training \\
        \addlinespace[1pt]
        Inference and Acceleration & Execute trained models efficiently through inference runtimes, cache and memory control, parallelism, kernels, and performance measurement. & Inference Runtime \\
         & & KV Cache/Memory \\
         & & Inference Parallelism \\
         & & Quantized Kernels \\
         & & Performance Benchmarking \\
        \addlinespace[1pt]
        Serving and Deployment & Expose models or AI workflows through APIs, deployable services, routing, authentication, and runtime boundaries. & Serving API \\
         & & Deployment Assets \\
         & & Autoscaling/Routing \\
         & & Authentication/Rate Limiting \\
         & & Multi Tenant Isolation \\
        \addlinespace[1pt]
        RAG and Retrieval & Build retrieval pipelines that load documents, create embedding indexes, retrieve and rerank context, and attach citations. & Document Loaders/Chunking \\
         & & Embedding/Indexing \\
         & & Retrieval/Reranking \\
         & & Citation Attribution \\
         & & Hybrid Search \\
        \addlinespace[1pt]
        Agents and Tooling & Implement agent control loops, tool or function calling, planning orchestration, memory state, and plugin integrations. & Tool/Function Calling \\
         & & Planning/Orchestration \\
         & & Memory State \\
         & & Integrations/Plugins \\
        \addlinespace[1pt]
        Evaluation and Benchmarking & Measure quality, safety, performance, and regression behavior for models, pipelines, or applications. & Quality Evaluation \\
         & & Safety Evaluation \\
         & & Performance Evaluation \\
         & & Regression Tests \\
        \addlinespace[1pt]
        Observability and LLMOps & Track experiments, register models, collect traces, metrics, and logs, and support CI/CD or governance workflows. & Experiment Tracking \\
         & & Model Registry \\
         & & Tracing/Metrics/Logs \\
         & & CI/CD Governance \\
        \addlinespace[1pt]
        UI and Workflows & Provide web interfaces, workflow builders, CLI/developer workflows, templates, and examples. & Web UI \\
         & & Workflow Builder \\
         & & CLI/Developer Workflows \\
         & & Templates/Examples \\
        \bottomrule
    \end{tabularx}
    \caption{AI infra module role taxonomy $\mathcal{V}_{\mathrm{role}}$ used for repository semantics.}
    \label{tab:module-role-summary}
\end{table*}
\clearpage

\end{document}